\documentclass[fleqn,usenatbib]{mnras}
\usepackage{mathtools}

\usepackage{breqn}
\usepackage[T1]{fontenc}
\usepackage{ae,aecompl}
\usepackage{ragged2e}

\usepackage{graphicx}	
\usepackage{amsmath}	
\usepackage{amssymb}	
\usepackage{subfig}
\usepackage{float}
\usepackage{multirow}
\usepackage{units}
\usepackage{placeins}

\usepackage{pdflscape}
\usepackage{rotating} %
\usepackage{xcolor}
\usepackage{epstopdf}
\usepackage{comment}

\definecolor{silver}{rgb}{0.75, 0.75, 0.75}

\title[Accretion scenario of BH-ULXs]{{Unveiling the accretion scenario of BH-ULXs using {\it XMM-Newton} observations}}

\author[Majumder et al.]{Seshadri Majumder$^{1}$\thanks{E-mail: smajumder@iitg.ac.in},
	Santabrata Das$^{1}$\thanks{E-mail: sbdas@iitg.ac.in}, 
	Vivek. K. Agrawal$^{2}$,
	Anuj Nandi$^{2}$\thanks{E-mail: anuj@ursc.gov.in} \\
	$^{1}$Department of Physics, Indian Institute of Technology Guwahati, Guwahati, 781039, India.\\
	$^{2}$Space Astronomy Group, ISITE Campus, U. R. Rao Satellite Centre, Outer
	Ring Road, Marathahalli, Bangalore, 560037, India.
}

\date{Accepted XXX. Received YYY; in original form ZZZ}

\pubyear{2022}

\begin{document}
	\label{firstpage}
	\pagerange{\pageref{firstpage}--\pageref{lastpage}}
	\maketitle
	
\begin{abstract}
		We present a comprehensive spectro-temporal analysis of five ultraluminous X-ray sources (ULXs) with central object likely being a black hole, using archival {\it XMM-Newton} observations. These sources, namely NGC 1313 X$-$1, NGC 5408 X-1, NGC 6946 X$-$1, M82 X$-$1 and IC 342 X$-$1, reveal short-term variability with fractional variance of $1.42-27.28\%$ and exhibit Quasi-periodic Oscillations (QPOs) with frequency $\nu_{\rm QPO} \sim 8-667$ mHz. Long-term evolution of ULXs energy spectra ($0.3 - 10$ keV; excluding M82 X$-$1) are described satisfactorily with a model combination that comprises a thermal Comptonization component (\texttt{nthComp}, yielding $\Gamma_{\rm nth} \sim 1.48 - 2.65$, $kT_{\rm e} \sim 1.62 - 3.76$ keV, $\tau \sim 8 - 20$, y-par $\sim 1.16 - 6.24$) along with a standard disc component (\texttt{diskbb}, $kT_{\rm in} \sim 0.16 - 0.54$ keV). We find that these ULXs generally demonstrate anti-correlation between disc luminosity and inner disc temperature as $L_{\rm disc} \propto T_{\rm in}^\alpha$, where $\alpha = - 3.58 \pm 0.04$ for NGC 1313 X$-$1 and IC 342 X$-$1, $\alpha = - 8.93 \pm 0.11$ for NGC 6946 X$-$1, and $\alpha = - 10.31 \pm 0.10$ for NGC 5408 X$-$1. We also obtain a linear correlation between bolometric luminosity $L_{\rm bol}$ and $\Gamma_{\rm nth}$ that indicates spectral softening of the sources when $L_{\rm bol}$ increases. We observe that in presence of QPO, Comptonized seed photon fraction varies in between $\sim 5 - 20 \%$, while the Comptonized flux contribution ($50 - 90\%$) dominates over disc flux. Utilizing $\nu_{\rm QPO}$ and $L_{\rm bol}$, we constrain ULXs mass by varying their spin ($a_{\rm k}$) and accretion rate ($\dot m$). We find that NGC 6946 X$-$1 and NGC 5408 X$-$1 seem to accrete at sub-Eddington accretion rate provided their central sources are rapidly rotating, whereas IC 342 X$-$1 and NGC 1313 X$-$1 can accrete in sub/super-Eddington limit irrespective to their spin values.
		
\end{abstract}
	


\section{Introduction}
\label{s:intro}

Ultraluminous X-ray sources (ULXs) are the bright off-nuclear point objects having luminosity in excess of Eddington luminosity ($\sim 10^{39}$ erg $\rm s^{-1}$ for $10 {\rm M}_\odot$ object) \cite[]{Fabbiano-etal1989,Makishima-etal20}. In general, the luminosities of these objects are found to lie in the range of $10^{39}-10^{41}$ erg $\rm s^{-1}$ \cite[]{Fabbiano-etal1989,Feng-etal2011}. Interestingly, the exact nature of the central accretor in ULXs as well as the plausible mechanisms to produce such a high luminosity still remain elusive \cite[for a recent review]{Fabrika-etal2021}. Meanwhile, several attempts were made to infer the characteristics of the ULXs. In an earlier attempt, it was indicated that the ULXs possibly harbor the stellar mass black hole (StMBH) that presumably accretes at super-Eddington rate \cite[and references therein]{Fabrika-etal2001, Watarai-etal2001, Ebisawa-etal2003, Poutanen-etal2007}. The alternative explanation rely on the fact that sub-Eddington accretion onto the intermediate mass black holes (IMBHs) of mass $10^{3}-10^{5}$$M_\odot$ seems to power the ULXs \cite[]{Colbert-etal1999, Makishima-etal2000, Miller-etal2003}. In addition, the possibility of harboring magnetized neutron star in ULXs is discussed to explain the discovery of coherent pulsation in M82 X$-$2 \cite[]{Bachetti-etal2014} followed by other such sources \cite[and references therein]{Furst-etal2016, Israel-etal2017, Carpano-etal2018, Sathyaprakash-etal2019, Rodriguez-etal2020, Quintin-etal2021}. An alternative scenario based on the sub-Eddington accretion onto StMBHs with beamed emission is also propounded \cite[]{Reynolds-etal1997, King-etal2002}, however it suffers lack of observational support \cite[]{Feng-etal2011}. Needless to mention that all the above efforts fail to converge in explaining the nature of the central accreter as well as the environment of the ULXs and hence remain under debate.

Interestingly, in the quest of ULXs, the study of the short-term variability including Quasi-periodic Oscillations (QPOs) observed in the power density spectra (PDS) are believed to be a viable tool as it carries the imprint signatures of the accretion dynamics and the nature of the accreter. So far, a handful ULXs are found to exhibit QPOs, namely NGC 1313 X$-$1 \cite[QPO frequency $\nu_{\rm QPO} \sim 290$ mHz,][]{Pasham-etal2015}; \cite[$\sim 80$ mHz, $\sim 268$ mHz and $\sim 297$ mHz,][]{Atapin-etal2019}, NGC 5408 X$-$1 \cite[$\sim 20$ mHz,][]{Strohmayer-etal2007};  \cite[$10-42$ mHz,][]{Strohmayer-etal2010,Atapin-etal2019}, M82 X$-$1 \cite[$\sim 54$ mHz,][]{Strohmayer-etal2003}; \cite[$\sim 110$ mHz,][]{Dewangan-etal2006}; \cite[$\sim 46$ mHz,][]{Caballero-etal2013}; \cite[$\sim 35$ mHz and $\sim 47$ mHz,][]{Atapin-etal2019}, NGC 6946 X$-$1 \cite[$\sim 8.5$ mHz,][]{Rao-etal2010}; \cite[$\sim 9.4$ mHz, $\sim 27$ mHz and $\sim 42$ mHz,]{Atapin-etal2019}, IC 342 X$-$1 \cite[$\sim 642$ mHz,][]{Agrawal-Nandi2015}. It may be noted that the characteristics of mHz QPOs resemble with the Type-C low frequency QPOs (LFQPOs) frequently observed in black hole X-ray binaries (BH-XRBs) \cite[and references therein]{Belloni-etal2005,Nandi-etal2012,Sreehari-etal2019}. Several studies indicate that the mass of ULXs can be determined from the appropriate scale factor corresponding to the QPO frequencies observed in BH-XRBs and ULXs, and eventually favors the existence of IMBH \cite[]{Dewangan-etal2006,Strohmayer-etal2009, Pasham-etal2012, Feng-etal2010, Agrawal-Nandi2015}. Further, using {\it RXTE} observations of M82 X$-$1, \cite{Pasham-etal2014} discovered a pair of coherent QPO frequency $\sim 3.32$ Hz and $\sim 5.07$ Hz with the harmonic ratio $3:2$. These QPOs can be interpreted as an analogous to the high frequency QPO (HFQPO) often seen in several BH-XRBs \cite[]{Belloni-etal2012,Sreehari-etal2020,Majumder-etal2022} and consequently, a $\sim 428 M_\odot$ black hole is expected to be present at the core of M82 X$-$1 \cite[]{Pasham-etal2014, Pasham-etal2015, Kaaret-etal2017}. Needless to mention that the estimation of source mass by means of QPO frequency is not always free from caveats \cite[]{Middleton-etal2011}.

The study of ULXs spectra is carried out over the decades starting from {\it ASCA} mission (energy range below $\sim 10$ keV) in late 1990s \cite[]{Colbert-etal1999}. Later on, rigorous spectral studies were performed using the high quality spectra from {\it XMM-Newton}, {\it Chandra} and {\it NuSTAR} observatories \cite[and references therein]{Walton-etal2018a, Walton-etal2019, Walton-etal2020, Gurpide-etal2021, Salvaggio-etal2022}. In general, ULXs spectra show two common features: a spectral turnover below $\sim 10$ keV and a soft excess below $\sim 2$ keV \cite[]{Stobbart-etal2006, Feng-etal2011, Ghosh-etal2022}. Based on the observed spectral morphologies, the ULXs spectra are classified into three spectral regimes, namely the {\it broadened disc, hard-ultraluminous} state and {\it soft-ultraluminous} state \cite[]{Gladstone-etal2009,Sutton-etal2013}. Meanwhile, spectral fitting of many ULXs with the empirical model combinations consisting of a standard accretion disc component and a Comptonization component indicates the presence of a cool inner accretion disc ($kT_{\rm in} \sim 0.1-0.3$ keV) and an optically thick ($kT_{e} \sim 1-2$ keV and $\tau > 6$ ) corona \cite[]{Gladstone-etal2009}. The inferred inner disc temperature implies a compact object of mass $\sim 10^{3} M_\odot$ (following $T_{\rm in}\propto M^{-0.25}$, \cite[]{Kaaret-etal2017}) that supports the IMBH interpretation with sub-Eddington accretion \cite[]{Feng-etal2011}. On the contrary, several ULXs are found to comply steep power-law ($\Gamma \gtrsim 2.5$) dominated spectral states of StMBH accreting at near-Eddington rate \cite[]{Winter-etal2006, Feng-etal2005,Feng-etal2011}. 

Models of accretion disk have been adopted to delineate the ULX spectra where two disc blackbody like components with different temperatures were used \cite[]{Stobbart-etal2006,Poutanen-etal2007,Kara-etal2020}. In these studies, it is suggested that the hotter component originates from the inner accretion flow by means of the inverse Comptonization of the soft photons by the swarm of hot electrons present in the corona \cite[]{Roberts-etal2007,Middleton-etal2015,Mukherjee-etal2015,Jithesh-etal2017}. On the contrary, the emission of the soft component is possibly emerged from the optically thick outflow originated within/from the spherization radius \cite[]{Shakura-etal1973, Poutanen-etal2007}. In addition, the slim disc model \cite[]{Abramowicz-etal1988} with varying radial temperature index is also found to be successful in describing the ULXs spectra \cite[see][]{Watarai-etal2001, Miyawaki-etal2006, Vierdayanti-etal2006,Ghosh-etal2021}. However, \cite{Gladstone-etal2009} indicate that the slim disk model perhaps inadequate in describing the spectra of ULXs resulting unphysical disc temperature ($kT_{\rm in} \gtrsim 3$ keV). 

Indeed, some of the ULXs may harbour neutron star accretor although they do not display any pulsed emission \cite[]{King-etal2017,Walton-etal2018b}. Meanwhile, piece-wise efforts were given in studying the temporal and spectral properties of the ULXs considering black hole accretor (hereafter BH-ULX) at its center \cite[]{DeMarco-etal2013,Luangtip-etal2021,Rao-etal2010,Stuchlik-Kolos2015,Brightman-etal2016b,Marlowe-etal2014,Agrawal-Nandi2015,Bachetti-etal2013}. Interestingly, some of the BH-ULXs are seen to exhibit QPO features, although their physical origin still remains elusive. Indeed, there are few occasions where combined spectro-temporal properties are studied to examine the characteristics of the BH-ULXs. \cite{Pasham-etal2012} found nearly non-varying spectral properties in most of the observations of NGC 5408 X$-$1, that evidently suggest the absence of its correlation with the observed QPO frequency. Further, despite having the lack of statistical significance, \cite{Caballero-etal2013} reported the decrease of $\nu_{QPO}$ with the increase of Comptonized flux for M82 X$-1$, however, the origin of such correlation is poorly understood.

Considering all these, in the present work, we aim to address the scarcity in understanding the accretion scenario of the BH-ULXs and perform the in-depth spectro-temporal analysis for these sources. Towards this, we revisit the archival {\it XMM-Newton} observations of the BH-ULXs that predominantly exhibit QPO features. While doing so, we investigate the fractional variability ($F_{\rm var}$) as well as the hardness intensity diagram (HID) in order to explain the variability properties of the source. We perform spectral modelling using the combined {\it XMM-Newton} data of {\it EPIC-PN} and {\it EPIC-MOS} instruments in $0.3-10$ keV energy range with both phenomenological and physical models. We observe significant spectro-temporal correlations for four BH-ULXs under consideration. Further, following a novel formalism developed by \cite{Das-etal2021}, we estimate the possible mass ($M_{\rm BH}$) range of these BH-ULX sources by means of their spin ($a_{\rm k}$) and accretion rates ($\dot m$).

The paper is organized as follows. In \S 2, we present the selection criteria of the BH-ULXs and briefly mention the source characteristics. In \S 3, we discuss the data reduction procedure of {\it EPIC-PN} and {\it EPIC-MOS} instruments. We present the results from timing and spectral analyses in \S 4 and \S 5, respectively. The spectro-temporal correlation results are studied in \S 6. In \S 7, we estimate the possible mass range of the BH-ULXs. Finally, we discuss the results and present conclusion in \S 8.

\begin{table*}
	\caption{Observation details of the five ULXs observed with {\it XMM-Newton}. In the table, source name, ObsID, Epoch, date of observation along with MJD and exposure time are mentioned. The count rate in $0.3-10$ keV energy range of the {\it EPIC-PN} detector along with hardness ratio (HR) and fractional variability ($F_{\rm var}$) in percentage are also tabulated. 	See text for details.}
	
	\renewcommand{\arraystretch}{1.2}
	\resizebox{1.0\textwidth}{!}{%
		\begin{tabular}{l l c c c c c c c c c c c c c c c}
			\hline\hline
			\multirow{2}{*}{Source} & \multirow{2}{*}{ObsID} & \multirow{2}{*}{Epoch$^{\otimes}$} & \multirow{2}{*}{Date} & \multirow{2}{*}{MJD} & \multirow{2}{*}{Effective} & \multirow{2}{*}{Rate (cts/s)} & \multirow{2}{*}{HR} & \multirow{2}{*}{$(F_{\rm var})^{\boxtimes}$} & \\ \\
			& & & & & Exposure (ks)& ($0.3-10$ keV)& ($\rm C_{2}/C_{1}$)$^*$& (\%) &\\
			\hline
			
			
			NGC 1313 X-1 & 0106860101$^{\dagger}$ & XM1 & 2000-10-17 & 51834.14 & 10.6 & $0.85\pm0.08$ & $0.38\pm0.08$ & $3.63\pm1.16$ \\
			
			& 0205230601$^{\dagger}$ & XM2  & 2005-02-07 & 53408.48 & 8.2 & $1.01\pm0.10$ & $0.37\pm0.09$ & $3.30\pm1.43$ \\

			& 0405090101 & XM3  & 2006-10-15 & 54023.98 & 38.0 & $0.85\pm0.08$ & $0.39\pm0.08$ & $3.48\pm2.27$ \\	
			
			& 0693850501 & XM4  & 2012-12-16 & 56277.67 & 32.4 & $1.03\pm0.08$ & $0.36\pm0.07$ & $3.36\pm1.79$ \\
			
			& 0693851201 & XM5  & 2012-12-22 & 56283.66 & 63.7 & $1.08\pm0.05$ & $0.27\pm0.05$ & $1.54\pm0.15$ \\
			
			& 0742590301$^{\dagger}$ & XM6 & 2014-07-05 & 56843.97 & 60.0 & $1.81\pm0.11$ & $0.34\pm0.05$ & $4.17\pm1.04$ \\
			
			& 0742490101$^{\dagger}$ & XM7 & 2015-03-30 & 57111.19 & 63.8 & $0.86\pm0.09$ & $0.40\pm0.09$ & $2.43\pm0.47$ \\
			
			& 0803990101$^{\dagger}$ & XM8 & 2017-06-14 & 57918.89 & 132.2 & $1.70\pm0.11$ & $0.36\pm0.05$ & $27.28\pm0.26$ \\
			
			& 0803990301$^{\dagger}$ & XM9 & 2017-08-31 & 57996.68 & 33.8 & $0.93\pm0.08$ & $0.39\pm0.08$ & $2.75\pm2.57$ \\
			
			& 0803990601$^{\dagger}$ & XM10 & 2017-12-09 & 58096.46 & 33.6 & $1.73\pm0.11$ & $0.33\pm0.05$ & $2.84\pm1.29$ \\
			\hline
			
			
			NGC 5408 X-1 & 0112290501$^{\dagger}$ & XM1 & 2001-07-31 & 52121.54 & 3.0 & $1.53\pm0.11$ & $0.05\pm0.02$ & $7.63\pm0.19$ \\
			
			& 0112291001$^{\dagger}$ & XM2 & 2002-07-29 & 52484.35 & 6.0 & $1.19\pm0.12$ & $0.15\pm0.05$ & $12.09\pm0.35$ \\
			
			& 0112291201$^{\dagger}$ & XM3 & 2003-01-27 & 52667.01 & 2.8 & $0.95\pm0.08$ & $0.08\pm0.03$ & $16.23\pm0.12$ \\
			
			& 0302900101 & XM4 & 2006-01-13 & 53748.77 & 81.2 & $1.08\pm0.05$ & $0.07\pm0.02$ & $9.12\pm1.01$ \\
			
			& 0500750101 & XM5 & 2008-01-13 & 54478.79 & 34.1 & $0.99\pm0.05$ & $0.08\pm0.03$ & $12.35\pm0.53$ \\
			
			& 0653380201 & XM6 & 2010-07-17 & 55394.13 & 77.3 & $1.22\pm0.06$ & $0.08\pm0.03$ & $7.29\pm1.11$ \\
			
			& 0653380301 & XM7 & 2010-07-19 & 55396.13 & 89.4 & $1.20\pm0.06$ & $0.08\pm0.02$ & $6.12\pm1.39$ \\
			
			& 0653380401 & XM8 & 2011-01-26 & 55587.67 & 80.5 & $1.14\pm0.06$ & $0.09\pm0.02$ & $6.87\pm1.31$ \\
			
			& 0653380501 & XM9 & 2011-01-28 & 55589.65 & 46.5 & $1.08\pm0.05$ & $0.09\pm0.03$ & $7.45\pm0.94$ \\
			
			& 0723130301 & XM10 & 2014-02-11 & 56699.02 & 18.7 & $0.99\pm0.02$ & $0.10\pm0.03$ & $10.33\pm0.46$ \\
			
			& 0723130401 & XM11 & 2014-02-13 & 56701.02 & 33.1 & $1.08\pm0.08$ & $0.10\pm0.02$ & $11.02\pm0.52$ \\
			\hline
			
			NGC 6946 X-1 & 0200670301 & XM1 & 2004-06-13 & 53169.78 & 9.7 & $0.39\pm0.05$ & $0.16\pm0.06$ & $14.61\pm0.65$ \\
			
			& 0500730201 & XM2 & 2007-11-02 & 54406.91 & 31.6 & $0.37\pm0.06$ & $0.17\pm0.07$ & $21.92\pm0.91$ \\
			
			& 0500730101 & XM3 & 2007-11-08 & 54412.94 & 18.5 & $0.38\pm0.06$ & $0.18\pm0.08$ & $26.96\pm0.74$ \\
			
			& 0691570101 & XM4 & 2012-10-21 & 56221.74 & 56.8 & $0.44\pm0.06$ & $0.19\pm0.07$ & $7.35\pm2.59$ \\
			
			& 0870830101$^{\dagger}$ & XM5 & 2020-07-08 & 59038.93 & $12.6$ & $0.42\pm0.08$ & $0.18\pm0.05$ & $24.52\pm2.26$ \\
			
			& 0870830201$^{\dagger}$ & XM6 & 2020-12-15 & 59198.32 & $16.1$ & $0.37\pm0.11$ & $0.19\pm0.06$ & $7.10\pm1.22$ \\
			
			& 0870830301$^{\dagger}$ & XM7 & 2021-04-04 & 59308.22 & $10.7$ & $0.38\pm0.06$ & $0.19\pm0.08$ & $13.30\pm0.89$ \\
			
			& 0870830401$^{\dagger}$ & XM8 & 2021-05-25 & 59359.89 & 12.5 & $0.34\pm0.05$ & $0.19\pm0.09$ & $23.82\pm0.55$ \\
			
			\hline
			
			M82 X-1 & 0112290201 & XM1 & 2001-05-06 & 52035.38 & 15.2 & $4.17\pm0.18$ & $1.08\pm0.09$ & $2.99\pm0.37$ \\
			
			& 0206080101 & XM2 & 2004-04-21 & 53116.90 & 29.0 & $3.17\pm0.16$ & $0.65\pm0.07$ & $1.42\pm0.48$ \\
			
			& 0560590101$^{\dagger}$ & XM3 & 2008-10-03 & 54742.88 & 19.2 & $8.01\pm0.27$ & $1.40\pm0.10$ & $8.87\pm0.16$ \\
			
			& 0560590301$^{\dagger}$ & XM4 & 2009-04-29 & 54950.36 & 15.2 & $4.91\pm0.20$ & $0.90\pm0.07$ & $2.51\pm0.40$ \\
			
			& 0657800101 & XM5 & 2011-03-18 & 55638.69 & 9.2 & $3.24\pm0.22$ & $0.83\pm0.05$ & $8.77\pm0.87$ \\
			
			& 0657801901 & XM6 & 2011-04-29 & 55680.55 & 3.9 & $3.05\pm0.15$ & $0.59\pm0.06$ & $3.03\pm0.26$ \\
			
			& 0657802101 & XM7  & 2011-09-24 & 55828.21 & 10.5 & $3.19\pm0.16$ & $1.00\pm0.09$  & $3.10\pm0.41$ \\
			
			\hline
			IC 342 X-1 & 0093640901$^{\dagger}$ &  XM1 & 2001-02-11 & 51951.06 & 5.4 & $0.47\pm0.06$ & $1.03\pm0.30$ & $3.44\pm1.89$ \\
			
			& 0206890101$^{\dagger}$ & XM2  & 2004-02-20 & 53055.31 & 6.6 & $0.98\pm0.09$ & $0.94\pm0.18$ & $2.45\pm1.49$ \\
			
			& 0206890201$^{\dagger}$ & XM3  & 2004-08-17 & 53234.82 & 19.6 & $0.55\pm0.07$ & $0.94\pm0.24$ & $6.42\pm1.64$ \\
			
			& 0693850601 &  XM4 & 2012-08-11 & 56150.83 & 36.0 & $0.52\pm0.06$ & $0.97\pm0.24$ & $3.85\pm3.58$ \\
			
			& 0693851301$^{\dagger}$ &  XM5 & 2012-08-17 & 56156.84 & 37.4 & $0.66\pm0.07$ & $0.91\pm0.19$ & $2.27\pm4.39$ \\
			
			\hline\hline
		\end{tabular}%
	}
	\label{table: Obs_details}
	\begin{list}{}{}
	\item $^*$$\rm C_{1}$ and $\rm C_{2}$ are the count rates in $0.3-2$ keV and $2-10$ keV energy ranges, respectively.
	\item $^{\dagger}$Non-detection of QPO.
	\item $^{\boxtimes}$Fractional variability, calculated from the background subtracted {\it EPIC-PN} light curves in $0.3-10$ keV energy range.
	\item $^{\otimes}$XM$i$ stands for XMM-Newton, where $i$ can run from $1$ to $11$ based on the observations.
\end{list}
\end{table*}

\section{Source selection and Observation}
\label{s:source}

In this work, we analyze the {\it XMM-Newton} \cite[]{Jansen-etal2001} observations of the ULXs that exhibit QPO features. Accordingly, we select five ULXs, namely NGC 1313 X$-$1, NGC 5408 X$-$1, NGC 6946 X$-$1, M82 X$-$1 and IC 342 X$-$1, and use the archival data available in \texttt{HEASARC} public data base\footnote{\url{https://heasarc.gsfc.nasa.gov/db-perl/W3Browse/w3browse.pl}}. Here, we consider the ULXs harbouring black hole (BH-ULXs) at their central core and carry out the analyses.

We consider all observations of the chosen sources for examining the correlations among the spectro-temporal properties. Note that these sources were observed for multiple times in a given year, although they do not display significant spectro-temporal changes in their characteristics. Hence, we consider only those observations which are typically separated by about six months interval to avoid repetition of results. Further, we consider three observations for NGC 1313 X$-$1 as it was observed six times in 2017 \cite[]{Walton-etal2020}. All the selected observations under consideration are listed in Table \ref{table: Obs_details}. We present brief description of the selected sources below.

\subsection{NGC 1313 X$-$1}

NGC 1313 X$-$1 is one of the three bright ULXs located in the nearby ($d \sim 4.25$ Mpc) barred spiral galaxy NGC 1313 \cite[]{Tully-etal2016}. The source luminosity is observed as $\sim 10^{40}$ erg $\rm s^{-1}$. The source belongs to the hard-ultraluminous spectral state and exhibits significant spectral variability \cite[]{Pintore-etal2012}. It is suggested that the source is powered by the super-Eddington accretion onto stellar-mass black hole \cite[]{Gladstone-etal2009, Bachetti-etal2013}. On the contrary, the sub-Eddington accretion on to IMBH of mass $2524-6811 M_\odot$ is also believed to be an alternative possibility \cite[]{Huang-etal2019}. The spin ($a_{\rm k}$) of the source is reported as $0.960-0.998$ \cite[]{Caballero-etal2010}, indicating a maximally rotating BH-ULXs.

\subsection{NGC 5408 X$-$1}

NGC 5408 X$-$1 is a nearby ULX source at a distance of 4.8 Mpc, located at $\sim 20$ arcsec off from the center of the host galaxy NGC 5408 \cite[]{Luangtip-etal2021}. The observed peak luminosity is found as $\sim 10^{40}$ erg $\rm s^{-1}$. The source is a good candidate for the timing studies as it exhibits variabilities on different time scales ranging from few tens of seconds to few months \cite[]{Strohmayer-etal2009b, Heil-etal2010, DeMarco-etal2013, Hernandez-Garcia-etal2015}. Employing the scaling law of QPO frequency as well as the dynamical mass measurement, the mass of the central source is obtained within IMBH regime \cite[]{Strohmayer-etal2009, DeMarco-etal2013, Cseh-etal2013, Luangtip-etal2021}. Using reflection spectroscopy, \cite{Caballero-etal2010} reported the spin of the source as $0.70_{-0.05}^{+0.12}$.

\subsection{NGC 6946 X$-$1}

NGC 6946 X$-$1 is one of the four ULXs in the spiral galaxy NGC 6946 at a distance of $7.72$ Mpc \cite[]{Anand-etal2018}. This source is found to be one of the most variable ULXs that demonstrates short term variabilities with fractional rms amplitude reaching up to $\sim 60\%$ \cite[]{Rao-etal2010}. 
In addition, the existence of soft lag at the lower frequencies and hard lag at relatively higher frequencies are reported \citep{Hernandez-Garcia-etal2015}. Upon scaling the mHz QPO frequency detected for this source with the low frequency QPOs (LFQPOs) of the galactic BH-XRBs, the mass of NGC 6946 X$-1$ is revealed as $(1-4) \times 10^{3}$ $M_\odot$ \citep{Rao-etal2010}. It is noteworthy to mention that the spin of the central source in NGC 6946 X$-1$ remains inconclusive till date.

\subsection{M82 X$-$1}

The ``cigar galaxy'' commonly known as M82 resides at a distance of $3.63$ Mpc \cite[]{Middleton-etal2015} and hosts two well-known ULX sources M82 X$-$1 and X$-$2 separated by only 5 arcsec \cite[]{Brightman-etal2020} in the sky plane. M82 X$-$1, being the most luminous ULX of the host galaxy with peak luminosity reaching up to $\sim 10^{41}$ erg $\rm s^{-1}$, remains one of the most promising IMBH candidate over the years. The mass of M82 X$-$1 is predicted as $300-810 M_\odot$ from the mass-luminosity correlation \cite[]{Feng-etal2010}, although \cite{Pasham-etal2014} inferred its mass as $\sim 400 M_\odot$ from the scaling relation of $3:2$ twin-peaked QPO frequency. Further, \cite{Stuchlik-Kolos2015} reported the ranges of mass and spin of the source as $140 M_\odot < M_{\rm BH} < 660 M_\odot$ and $0.05 < a_{\rm k} < 0.6$, respectively. 

\subsection{IC 342 X$-$1}

IC 342 is a nearby spiral galaxy at a distance $\sim 3.61$ Mpc \cite[]{Middleton-etal2015} and it contains two ULXs, namely IC 342 X$-$1 and X$-$2. The source IC 342 X$-$1 is one of the most studied ULX sources since its discovery by \textit{Einstein} IPC observations \cite[]{Fabbiano-etal1987}. Subsequently, \textit{ROSAT} \cite[]{Bregman-etal1993, Roberts-etal2000} and \textit{ASCA} observations \cite[]{Kubota-etal2001} also confirmed its ultra-luminous state with maximum observed luminosity $\gtrsim 10^{39}$ erg $\rm s^{-1}$. The spectral state transitions, namely from the disc-dominated state to the power-law dominated state, are first observed in IC 342 X$-$1 \cite[]{Kubota-etal2001}. Several studies from observational and theoretical fronts predict the mass of IC 342 X$-1$ in the range of $20-1783$ $M_\odot$ \cite[]{Marlowe-etal2014, Agrawal-Nandi2015, Das-etal2021}. Moreover, the spinning nature of the source still remains inconclusive.

\section{Data reduction}

\label{s:Data-reduction}

We follow the standard data reduction procedure\footnote{\url{https://www.cosmos.esa.int/web/xmm-newton/sas-threads}} 
using the software \texttt{SCIENCE ANALYSIS SYSTEM (SAS) V19.1.0}\footnote{\url{https://www.cosmos.esa.int/web/xmm-newton/sas}} of {\it XMM-Newton} instrument. Note that the data of all the observations are available in {\it PrimeFullWindow} mode for the sources under consideration. For spectral analysis, we use combined {\it EPIC-PN} \cite[]{Struder-etal2001} and {\it EPIC-MOS} data, whereas the timing analysis is carried out using {\it EPIC-PN} data due to better timing resolution ($\sim 73.4$ ms) of the instrument. The event files are generated using the standard task \texttt{epchain} and \texttt{emproc} for {\it EPIC-PN} and {\it EPIC-MOS}, respectively. First, we generate a light curve in $10-15$ keV energy range to identify the effect of particle background flares in the observations. Following \cite{Agrawal-Nandi2015}, the Good Time Intervals (GTIs) are generated using the task \texttt{tabgtigen} by adopting the selection criteria \texttt{RATE} $\leq 3 \times$ mean count rate of the light curve in $10-15$ keV energy range. The above selection criteria removes all the background flares including data gaps in the light curves. We consider data corresponding to the longest continuous observation available in the GTI segments. Finally, the cleaned event file is generated using the selected GTI interval to extract the science products. The events are selected according to the selection criteria \texttt{PATTERN} $\leq 4$ and \texttt{PATTERN} $\leq 12$ with \texttt{FLAG}$=$0 for {\it EPIC-PN} and {\it EPIC-MOS}, respectively. The task \texttt{evselect} is used to generate light curves which are further corrected with background subtraction and various instrumental effects, such as vignetting, bad pixels using the tool \texttt{epiclccorr}. The source and background spectra are extracted using the task \texttt{evselect}. Instrument response file (rmf) and ancillary response file (arf) are generated using the tasks \texttt{rmfgen} and \texttt{arfgen}, respectively. 

Following \cite{Kajava-etal2009,Agrawal-Nandi2015}, light curves and spectra are extracted from a $40$ arcsec circular region concentric with the source coordinate for all the sources except M82 X$-$1 and NGC 6946 X$-$1. However, the source region is reduced to 30 arcsec in few of the observations including NGC 6946 X$-$1, provided the source falls near the chip gap of pn camera. A smaller circular source region of 18 arcsec is chosen for M82 X$-$1 to minimize the chance of contamination from nearby sources \cite[]{Dewangan-etal2006}. Backgrounds are estimated by extracting the counts from the identical circular region resides away from the source.

\begin{table*}
	\centering
	\caption{Details of the QPO characteristics obtained from the best fitted power density spectra of five BH-ULXs in $0.3-10$ keV energy range. Here, $\nu_{\rm QPO}$, FWHM, Q-factor and $\sigma$ denote the centroid frequency, width, quality factor and significance of the QPO features. $\rm QPO_{\rm rms}\%$ and $\rm Total_{\rm rms}\%$ denote the percentage rms amplitudes of the QPOs and the entire PDS, respectively. All the errors are computed with $68\%$ confidence level. See text for details.}
	
	\renewcommand{\arraystretch}{1.4}
	
	\resizebox{1.0\textwidth}{!}{%
		\begin{tabular}{l @{\hspace{0.4cm}} c @{\hspace{0.4cm}} c c @{\hspace{0.3cm}} c @{\hspace{0.3cm}} c @{\hspace{0.3cm}} c @{\hspace{0.5cm}} c @{\hspace{0.3cm}} c @{\hspace{0.4cm}} c @{\hspace{0.4cm}} c @{\hspace{0.4cm}} c @{\hspace{0.4cm}} c @{\hspace{0.01cm}} c }
			
			\hline\hline
			Source & ObsID & MJD & $\nu_{\rm QPO}$ & FWHM & Q-factor$^{\boxtimes}$ & Significance & $\rm QPO_{\rm rms}$ & $\rm Total_{\rm rms}$ & $\chi^{2}/dof$ & No. of Newbins & \\
			
			& & & (mHz) & (mHz) & & $(\sigma)$ & $(\%)$ & $(\%)$ & & per Interval &\\
			
			
			\hline
			NGC 1313 X-1 & 0405090101 & 54023.98 & $82.14_{-1.58}^{+1.35}$ & $9.12_{-6.77}^{+4.64}$ & 9.01 & 3.34 & $8.17\pm0.49$ & $22.59\pm1.27$ & 107/122 & 256 &\\
			
			& 0693850501 & 56277.67 & $304.60_{-4.99}^{+4.21}$ & $33.34_{-11.05}^{+11.12}$ & $9.14$ & $4.67$ & $11.29\pm0.59$ & $18.74\pm1.54$ & 16/36 & 512 &\\
			
			& & & $667.31_{-47.19}^{+37.60}$ & $68.63_{-22.58}^{+33.14}$ & $9.72$ & $1.62$ & $7.04\pm0.18$ & $18.74\pm1.54$ & 16/36 & 512 &\\
			
			& 0693851201 & 56283.66 & $305.04_{-3.58}^{+6.22}$ & $37.25_{-10.18}^{+15.23}$ & $8.18$ & $2.76$ & $11.46\pm0.72$ & $15.69 \pm 0.95$ & 51/50 & 256 &\\
			\hline
			
			
			NGC 5408 X-1 & 0302900101 & 53748.77 & $18.51_{-1.14}^{+1.06}$& $15.66_{-2.43}^{+2.83}$ & 1.18 & 9.41 & $15.37\pm1.24$ & $27.17\pm10.84$ & 542/506 & 1024 &\\
			
			& 0500750101 & 54478.79 & $9.02_{-0.71}^{+0.67}$ & $2.56_{-2.08}^{+2.45}$ & $3.52$ & $2.41$ & $7.81\pm2.62$ & $28.97\pm12.90$ & 1225/1018 & 2048 &\\
			
			& 0653380201 & 55394.13 & $38.13_{-0.50}^{+0.28}$ & $25.07_{-1.19}^{+1.41}$ & 1.52 & 4.02 & $12.27\pm1.19$ & $31.40\pm3.47$ & 210/247 & 512 &\\
			& &  & $97.94_{-2.72}^{+2.76}$ & $10.79_{-0.48}^{+0.74}$ & 9.08 & 2.73 & $6.16\pm0.48$ & $31.40\pm3.47$ & 210/247 & 512 &\\
			
			& 0653380301 & 55396.13 & $39.14_{-0.23}^{+0.27}$ & $6.54_{-0.50}^{+0.62}$ & 6.08 & 1.94 & $4.89\pm0.50$ & $23.54\pm2.46$ & 76/58 & 256 &\\
			
			& 0653380501 & 55589.65 & $17.99_{-1.93}^{+1.58}$ & $14.31_{-3.81}^{+4.41}$ & 1.26 & 5.40 & $14.34\pm1.81$ & $28.33\pm8.09$ & 505/506 & 1024 &\\
			
			& 0723130301 & 56699.02 & $11.09_{-1.04}^{+0.91}$ & $6.68_{-2.74}^{+3.36}$ & 1.66 & 3.51 & $13.92\pm4.23$ & $34.90\pm9.10$ & 33/40 & 256 &\\
			&  &  & $44.22_{-1.12}^{+1.84}$ & $3.87_{-2.79}^{+3.05}$ & 11.44 & 2.38 & $7.41\pm1.41$ & $34.90\pm9.10$ & 33/40 & 256 &\\
			
			& 0723130401 & 56701.02 & $13.08_{-0.85}^{+0.74}$ & $7.39_{-2.47}^{+3.44}$ & 1.77 & 4.48 & $14.42\pm3.98$ & $32.10\pm11.65$ & 566/506 & 1024 &\\
			\hline
			
			
			NGC 6946 X-1 & 0200670301 & 53169.78 & $38.71_{-2.53}^{+5.61}$ & $9.67_{-5.84}^{+6.22}$ & 4.00 & 2.04 & $29.05\pm0.51$ & $48.98\pm6.01$ & 149/122 & 512 &\\
			
			& 0500730201 & 54406.91 & $9.23_{-0.92}^{+0.81}$ & $5.85_{-2.02}^{+2.43}$ & 1.59 & 3.79 & $20.60\pm2.06$ & $44.18\pm12.32$ & 1168/1018 & 2048 &\\
			
			& 0500730101 & 54412.94 & $8.44_{-0.35}^{+0.43}$ & $1.91_{-1.12}^{+1.19}$ & 4.40 & 2.90 & $13.41\pm2.13$ & $38.35\pm8.58$ & 62/74 & 1024 &\\
			
			& 0691570101 & 56221.74 & $41.72_{-5.69}^{+5.04}$ & $41.12_{-13.09}^{+12.52}$ & 1.01 & 4.32 & $23.47\pm0.61$ & $41.13\pm2.74$ & 133/122 & 256 &\\
			
			\hline
			
			
			M82 X$-$1 $^\otimes$ & 0112290201 & 52035.38 & $55.85_{-2.01}^{+1.44}$ & $6.44_{-3.45}^{+5.89}$ & 8.67 & 2.37 & $11.54 \pm 4.33$ & $32.68\pm10.30$ & 637/506 & 2048 &\\
			
			& 0206080101 & 53116.90 & $110.81_{-4.26}^{+5.03}$ & $31.82_{-13.47}^{+19.72}$ & 3.48 & 3.61 & $12.13 \pm 2.18$ & $20.21\pm4.06$ & 80/105 & 512 &\\
			
			& 0657800101 & 55638.69 & $45.55_{-2.19}^{+1.61}$ &  $5.85_{-2.73}^{+4.41}$ & 7.78 & 2.21 & $21.88 \pm 7.75$ & $68.48\pm12.84$ & 635/504 & 2048 &\\
			
			& 0657801901 & 55680.55 & $47.63_{-3.22}^{+5.47}$ & $17.13_{-5.28}^{+6.32}$ & 2.78 & 2.94 & $32.95 \pm 11.18$ & $64.62\pm24.32$ & 362/250 & 512 &\\
			
			& 0657802101 & 55828.21 & $35.74_{-2.44}^{+2.53}$ & $11.48_{-3.73}^{+5.05}$ & 3.12 & 3.33 & $28.30 \pm 10.64$ & $42.74\pm14.28$ & 53/58 & 1024 &\\
			
			
			\hline
			IC 342 X-1 & 0693850601 & 56150.83 & $643.40_{-23.89}^{+15.30}$ & $112.20_{-46.52}^{+59.69}$ & 5.73 & 3.75 & $25.82\pm0.46$ & $48.30\pm13.74$ & 32/51& 256 &\\
			\hline
			\hline
			
		\end{tabular}%
	}
\begin{list}{}{}
	\item $^{\boxtimes}$$Q \lesssim 2$ indicates `QPO like' broad features \cite[][for detail]{van-der-Klis2006}.
	\item $^{\otimes}$Estimated ${\rm QPO}_{\rm rms}\%$ considering possible flux contamination due to M82 X$-$2 (see Appendix-A for details).
\end{list}

	\label{table:PDS_parameters}
\end{table*}

\section{Timing analysis and results}
\label{s:temporal}

Here, we explore the temporal variabilities of five BH-ULXs under considerations and the obtained results are presented in the sub-sequent sections. 

\subsection{Variability and Hardness-Intensity Diagram (HID)}

\begin{figure}
	\begin{center}
		\includegraphics[width=\columnwidth]{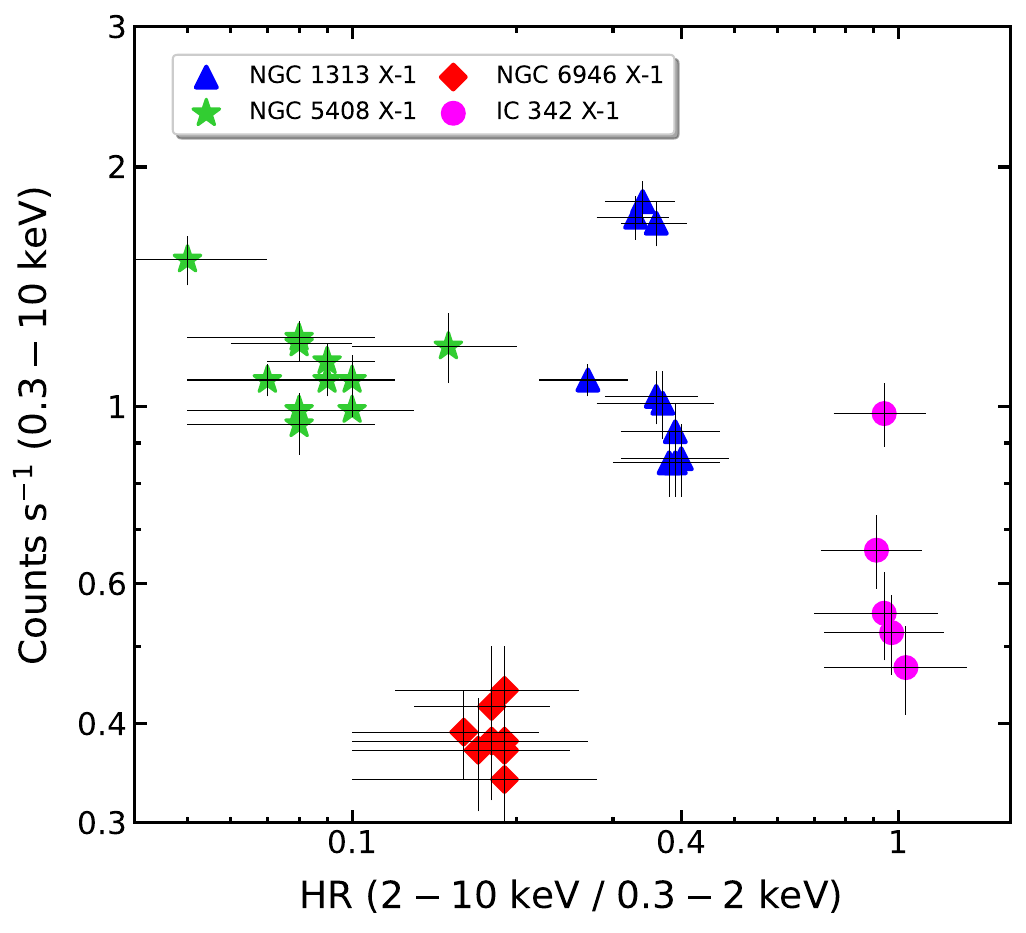}
	\end{center}
	\caption{Hardness-intensity diagram (HID) of the BH-ULXs. The variation of {\it EPIC-PN} background subtracted count rate in $0.3-10$ keV energy range with the hardness ratio is shown for each source. See text for details.}
	\label{fig:HID}
\end{figure}

\begin{figure}
	\begin{center}
		\includegraphics[width=\columnwidth]{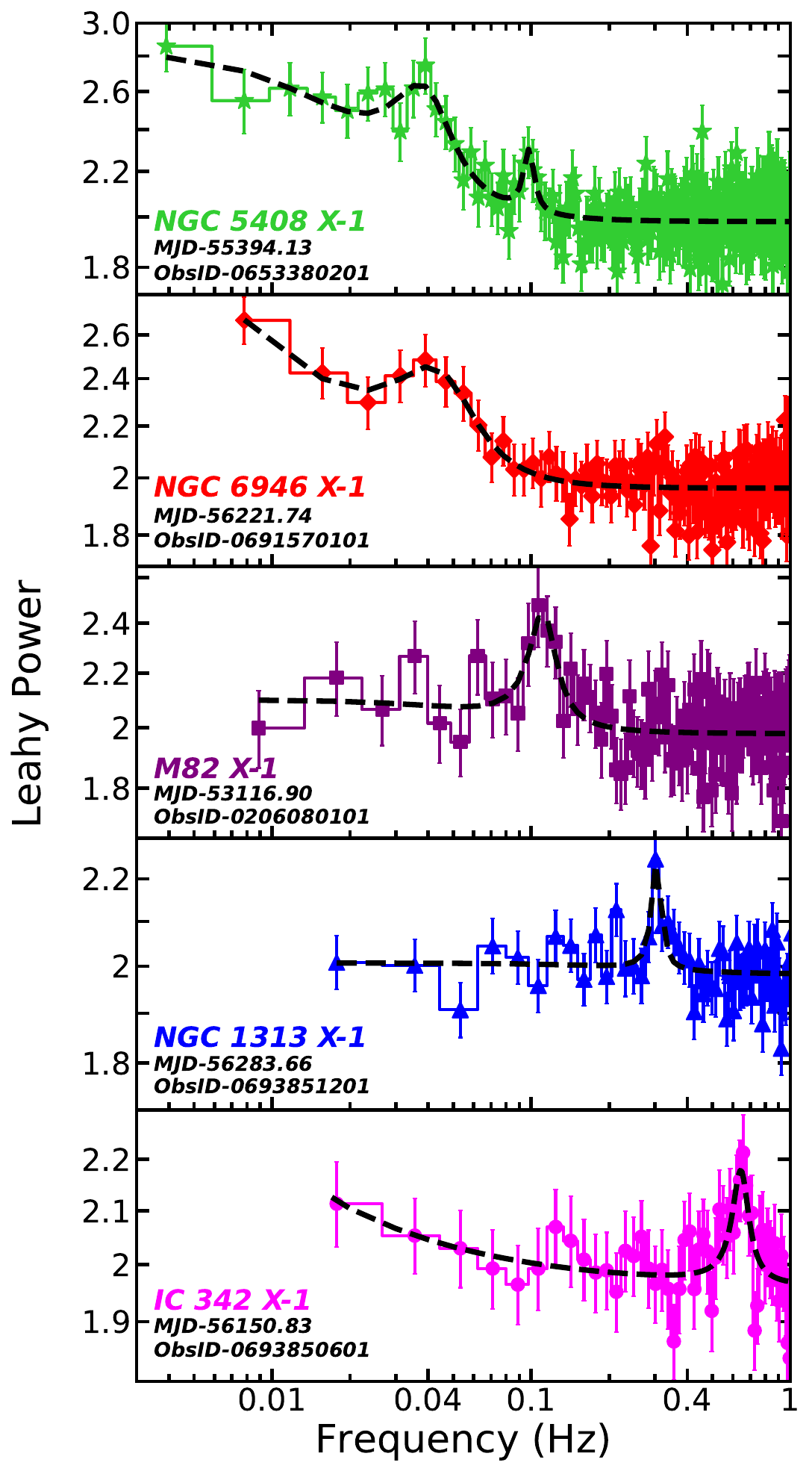}
	\end{center}
	\caption{Leahy normalized power density spectra obtained from {\it EPIC-PN} data of the five BH-ULXs. In each panel, observed highest QPO frequency in $0.3-10$ keV energy range for a given source is are presented along with source name and observation details. Each PDS is modelled with the combination of \texttt{Lorentzians} and a \texttt{constant}. See text for details.}
	\label{fig:PDS_QPO}
\end{figure}

We study the short-term variability of five BH-ULXs under consideration. While doing so, background subtracted $200$ s binned {\it EPIC-PN} light curves are generated for each of the sources in $0.3-10$ keV energy range. We compute the fractional variability ($F_{\rm var}$) {\it i.e.,} the rms variability amplitude normalized to mean count rate \cite[]{Edelson-etal2002, Vaughan-etal2003, Pintore-etal2014, Bhuvana-etal2021} in $0.3-10$ keV energy range to deduce the variability of the sources in the time domain. The fractional variabilities of all five sources are found to vary in the range of $1.42-27.28$ per cent, whereas NGC 6946 X$-$1 is appeared to be most variable with $F_{\rm var} \sim 27$ per cent. The mean count rate and the fractional variability corresponding to each observations are tabulated in Table \ref{table: Obs_details}.

Further, we study the hardness intensity diagram (HID) of the four BH-ULXs. The hardness ratio is defined as ${\rm HR}=\rm C2/C1$, where $\rm C1$ and $\rm C2$ are the background subtracted count rates in $0.3-2$ keV and $2-10$ keV energy bands, respectively. It may be noted that the soft energy band is considered up to $\sim 2$ keV as the disc emission generally dominates within this energy range for ULXs (see \S 5). The obtained HR values are mentioned in Table \ref{table: Obs_details} for all the observations. In Fig. \ref{fig:HID}, we present the HID of the four BH-ULXs, where source count rate is plotted with HR in logarithmic scales. In the HID plot, results corresponding to various sources are presented using different point styles as shown in the legend. We observe that the intensity of NGC 5408 X$-$1 noticeably varies with HR, although the change of HR seems to be marginal for other three sources.

\subsection{Power Spectral Properties}

We generate power density spectra (PDS) for all the observations using the {\it EPIC-PN} data in $0.3-10$ keV energy range. The ftools task  \texttt{powspec}\footnote{\url{https://heasarc.gsfc.nasa.gov/xanadu/xronos/examples/powspec.html}} available in \texttt{HEASOFT V6.29.1}\footnote{\url{https://heasarc.gsfc.nasa.gov/xanadu/xspec/}} is used to obtain the PDS. Background subtracted light curves of $0.5$ s time resolution are used to generate the PDS with Nyquist frequency of $1$ Hz. In general, the light curves are divided into shorter segments of equal duration to generate individual power spectrum. All the power spectra are further averaged in a single frame to obtain the resultant PDS. We notice that the number of newbins per interval (in \texttt{powspec}), which determines the duration of the light curve segments, plays an important role in order to maximize the signal to noise ratio in the power spectra. In particular, we observe that the specific choice of the number of newbins per interval results the prominent feature of individual QPOs provided the minimum frequency in the PDS is chosen as roughly one-tenth to one-thirtieth of the observed QPO frequency. A similar method is adopted in \cite{Atapin-etal2019} to obtain the power density spectra of the sources under consideration. We choose specific number of newbins per interval (see Table. \ref{table:PDS_parameters}) to construct the individual PDS, which is further averaged in a single frame to obtain the resultant PDS for the significant detection of QPO features. A constant binning factor of $1.04$ is used to rebin the final PDS in the frequency space. Note that the number of newbins per interval plays an important role in order to maximize the signal to noise ratio in the power spectra to detect the prominent QPO features. The number of newbins per interval is presented in Table \ref{table:PDS_parameters}.

Each PDS expressed in Leahy normalization \cite[]{Leahy-etal1983} represents a Poisson noise of power level $\sim 2$ with flat-topped noise (FTN) at the low frequency along with QPO features. Following \cite{Caballero-etal2013}, we adopt the model combination comprising of a \texttt{constant} (for the Poisson noise power), zero centroid \texttt{Lorentzian} (for the low frequency FTN) and an additional \texttt{Lorentzian} profile (for the QPO feature) to obtain the best fit power spectra. Each Lorentzian is characterized by three parameters, namely centroid (LC), width (LW) and normalization (LN) \cite[see][and references therein]{Sreehari-etal2020,Majumder-etal2022}. The model fitted PDS of five sources are depicted in the respective panels of Fig. \ref{fig:PDS_QPO}. 
In the top panel, we display the power spectrum of NGC 5408 X$-$1 (ObsID$-$0653380201), where the poisson noise level and the FTN are fitted with a \texttt{constant} and a zero centroid \texttt{Lorentzian} component. Next, we include one \texttt{Lorentzian} at $\sim 40$ mHz to fit the excess positive residuals near this frequency. The fit is obtained with a $\chi_{\rm red}^{2}$ of $219/250=0.88$, although significant positive residuals still remain around $\sim 100$ mHz. To address this, another \texttt{Lorentzian} is added to the composite model at around $\sim 100$ mHz. The resultant fit is obtained with a $\chi_{\rm red}^{2} = 0.85$. We follow the above methodology to fit all the remaining PDS corresponding to each source and find that the requirement of the number of model components generally vary with observations. It is perceived that the FTN in the PDS can be well-fitted using the zero centroid Lorentzian of all the sources except IC 342 X$-$1. A power-law profile with index $\sim 0.73$ provides a better fit of the low frequency FTN component of IC 342 X$-$1.

The best fit PDS of NGC 5408 X$-$1 (ObsID$-$0653380201) confirms simultaneous detection of two QPOs at $\sim 38.13_{-0.50}^{+0.28}$ mHz and $\sim 97.94_{-2.72}^{+2.76}$ mHz, respectively. The significance ($\sigma = LN/{\rm err}_{\rm neg}$, where ${\rm err}_{\rm neg}$ being the negative error of normalization of the fitted \texttt{Lorentzian}) of the detected QPOs are obtained as $4.02 \sigma$ and $2.73 \sigma$. The percentage rms amplitudes of $\sim 38.13$ mHz and $\sim 97.94$ mHz QPOs are estimated \cite[and references therein]{Sreehari-etal2020,Majumder-etal2022} as $12.27\pm1.19$ percent and $6.16\pm0.48$ percent, respectively. Further, we calculate the rms amplitudes of the entire PDS in $0.001-1$ Hz frequency range which is obtained as $31.40\pm3.47$ percent. Following the above approach, we find the presence of QPO features in twenty one observations for all BH-ULXs and two of them exhibit twin peak QPO features (for NGC 1313 X$-$1 and NGC 5408 X$-$1, the $\nu_{QPO} \sim 667$ mHz and $\sim 98$ mHz) that appear simultaneously in the PDS. The model fitted parameters and the estimated parameters of the QPO features are presented in Table \ref{table:PDS_parameters}. 

\section{Spectral analysis and results}

\label{s:spectral}

We generate {\it XMM-Newton} energy spectra for all the observations in $0.3-10$ keV energy range. \texttt{XSPEC V12.12.1f} \cite[]{Arnaud-etal1996} of \texttt{HEASOFT V6.29.1} is used to model the energy spectrum. While spectral fitting, we follow the standard procedure to extract the background spectrum file, instrument response file (rmf) and ancillary response file (arf). Each spectrum is grouped with $25$ counts per spectral bin using \texttt{specgroup} tool of \texttt{SAS V19.1.0}. We fit {\it EPIC-PN} and {\it EPIC-MOS} spectra simultaneously to improve the counting statistics. The cross-calibration between different instruments is taken care by including a \texttt{constant} component in the spectral fitting \cite[]{Pintore-etal2014}. For NGC 5408 X$-1$, we include a plasma component (\texttt{APEC} in \texttt{XSPEC}) of temperature $\sim 0.9$ keV to adjust the low energy residuals \cite[see][]{Pintore-etal2014}. Following \cite{Earnshaw-etal2019}, we include a \texttt{gaussian} at $\sim 1$ keV to fit the NGC 6946 X$-1$ spectra. We do not include M82 X$-$1 in the spectral study as the effect of contamination due to the close proximity (within 5 arcsec) of M82 X$-$2 can not be ruled out \cite[]{Brightman-etal2016a}.

\subsection{Modelling of {\it XMM-Newton} Spectra}

\begin{figure}
	\begin{center}
		\includegraphics[width=\columnwidth]{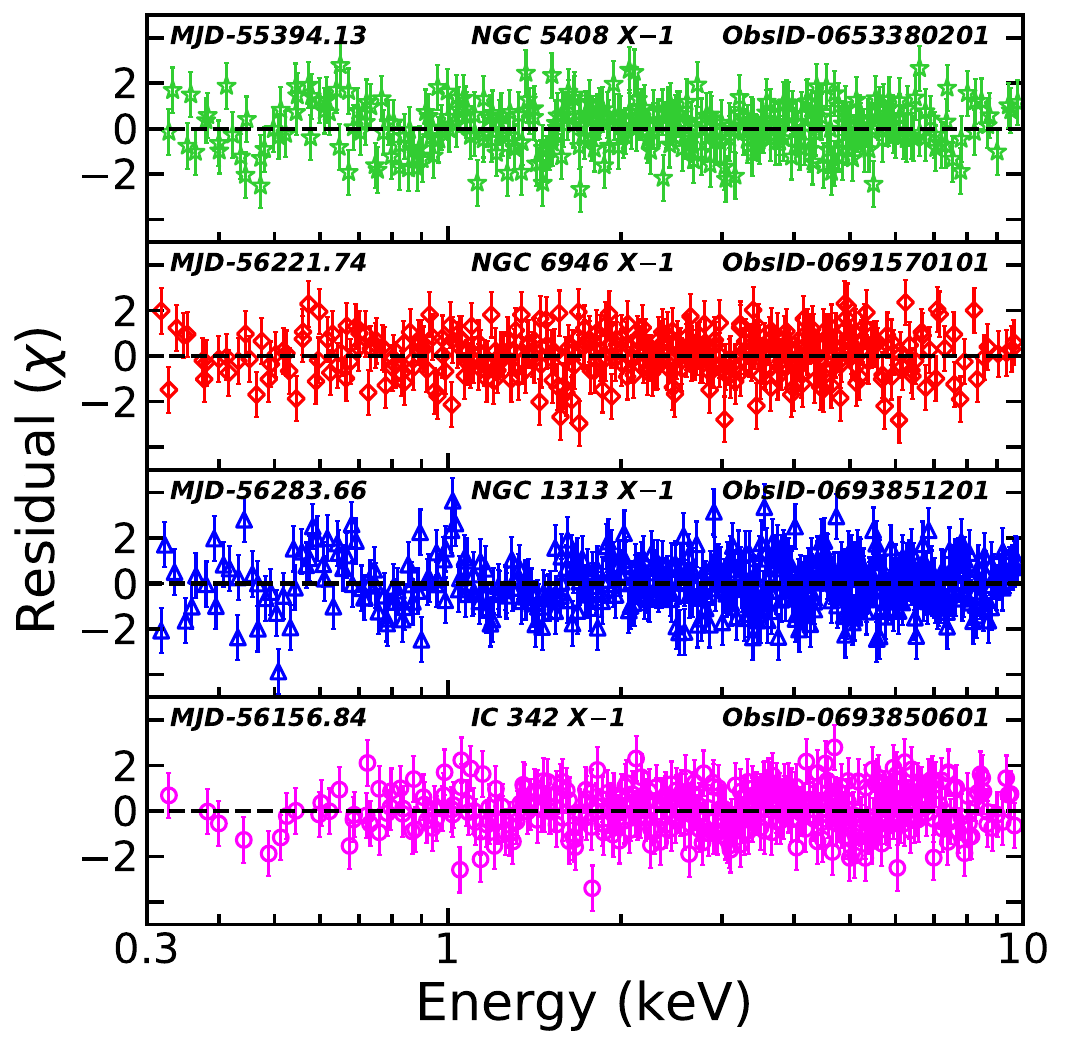}
	\end{center}
	\caption{Variation of the residuals in units of $\sigma$. In each panel, the residuals obtained from the model fitting of combined {\it EPIC-PN} and {\it EPIC-MOS} energy spectra with the model \texttt{TBabs}$\times$(\texttt{diskbb} + \texttt{nthComp}) are shown. See text for details.}
	\label{fig:residual_nth}
\end{figure}

\begin{table*}
	\caption{Details of the best fitted model parameters of combined {\it EPIC-PN} and {\it EPIC-MOS} spectra in $0.3-10$ keV energy range with the model \texttt{TBabs$\times$(diskbb + simpl$\times$diskbb)} for four ULXs as mentioned in Table \ref{table: Obs_details}. In the table, $n_{\rm H}$ is the hydrogen column density, $kT_{\rm in}^{\rm soft}$ is the temperature of the soft disc component in keV, $kT_{\rm in}^{\rm hard}$ is the temperature of relatively hotter disc component in keV. $\Gamma_{\rm simpl}$ and $f_{\rm scatt}$ is the power-index and Comptonized seed photon fraction of the \texttt{simpl} component, respectively.  $\chi^2/d.o.f$ and $P_{\rm null}$ refer the reduced $\chi^2$ and null hypothesis probability values. $F_{\rm disc}^{\rm soft}$ and $F_{\rm disc}^{\rm hard}$ are the flux values in erg cm$^{-2}$ s$^{-1}$ associated with hotter and softer \texttt{diskbb} component, respectively. $F_{\rm bol}$, and $L_{\rm bol}$ are the bolometric flux and luminosity, respectively. All the errors are computed with $90$ per cent confidence level. See text for details.}
	
	\renewcommand{\arraystretch}{1.6}
	
	\label{table:disc_spec_parameters}
	
	\resizebox{1.0\textwidth}{!}{%
		\begin{tabular}{l @{\hspace{0.2cm}} l @{\hspace{0.2cm}} c @{\hspace{0.2cm}} c @{\hspace{0.2cm}} c @{\hspace{0.4cm}} c @{\hspace{0.4cm}} c @{\hspace{0.4cm}} c @{\hspace{0.4cm}} c @{\hspace{0.5cm}} c @{\hspace{0.5cm}} c @{\hspace{0.001cm}} c @{\hspace{0.001cm}} c @{\hspace{0.001cm}} c @{\hspace{0.001cm}} c @{\hspace{0.001cm}} c @{\hspace{0.001cm}} c}
			\hline
			\hline
			& & & & \multicolumn{3}{|c|}{Model fitted parameters} & & \multicolumn{9}{|c|}{Estimated parameters} \\
			
			\cline{4-8}
			\cline{11-14}
			& & & & & & & & & & & & & & & & \\
			Source & ObsID & MJD & $n_{\rm H}$ & $kT_{\rm in}^{\rm soft}$ & $kT_{\rm in}^{\rm hard}$ & $\Gamma_{\rm simpl}$ & $f_{\rm scatt}$ & $\chi^2/d.o.f$ & $P_{\rm null}$$^{\boxtimes}$ & $F_{\rm disc}^{\rm soft}$ & $F_{\rm disc}^{\rm hard}$ & $F_{\rm bol}$ & $L_{\rm bol}$& \\

			& & & & & & & & & \multicolumn{3}{|c|}{($0.3-10$ keV)} & ($0.1-100$ keV) &  \\
			
			& & & ($10^{22}$ ${\rm cm}^{-2}$) & (keV) & (keV) & & (\%) & & & & \multicolumn{1}{|c|}{($\tiny{10^{-12}}$ erg ${\rm cm}^{-2}$ ${\rm s}^{-1}$)} & & ($10^{40}$ erg ${\rm s}^{-1}$) \\
			
			\hline
			
			
			NGC 1313 X-1 & 0106860101$^{\dagger}$ & 51834.14 & $0.21_{-0.03}^{+0.03}$ & $0.30_{-0.03}^{+0.04}$ & $2.35_{-0.21}^{+0.25}$ & $-$ & $-$ & $131/131$ & $0.47$ & $1.50_{-0.06}^{+0.06}$ & $2.32_{-0.09}^{+0.09}$ & $4.50_{-0.09}^{+0.09}$ & $0.97\pm0.01$ \\

			(d $=$ 4.25 ${\rm Mpc}^{c}$) & 0205230601$^{\dagger}$ & 53408.48 & $0.40_{-0.02}^{+0.02}$ & $0.19_{-0.03}^{+0.01}$ & $0.76_{-0.16}^{+0.03}$ & $1.09_{-0.01}^{+0.15}$ & $60.30_{-1.84}^{+1.86}$ & $217/205$ & $0.27$ & $5.01_{-0.19}^{+0.19}$ & $3.32_{-0.13}^{+0.13}$ & $12.69_{-0.32}^{+0.32}$ & $2.74\pm0.07$ \\

			& 0405090101 & 54023.98 & $0.30_{-0.01}^{+0.02}$ & $0.21_{-0.01}^{+0.01}$ & $2.07_{-0.17}^{+0.22}$ & $1.76_{-0.07}^{+0.08}$ & $14.20_{-1.25}^{+1.29}$ & $502/445$ & $0.03$ &  $2.36_{-0.02}^{+0.02}$ & $0.92_{-0.01}^{+0.01}$ & $6.63_{-0.04}^{+0.04}$ & $1.43\pm0.01$ \\
			
			& 0693850501 & 56277.67 & $0.27_{-0.02}^{+0.03}$ & $0.24_{-0.04}^{+0.02}$ & $2.38_{-0.44}^{+0.71}$ & $1.85_{-0.14}^{+0.15}$ & $19.70_{-1.82}^{+1.23}$ & $489/439$ & $0.05$ & $2.53_{-0.03}^{+0.03}$ & $1.00_{-0.03}^{+0.03}$ & $7.03_{-0.08}^{+0.08}$ & $1.52\pm0.02$ \\
			
			& 0693851201 & 56283.66 & $0.27_{-0.02}^{+0.02}$ & $0.23_{-0.02}^{+0.02}$ & $2.08_{-0.35}^{+0.46}$ & $1.83_{-0.09}^{+0.16}$ & $20.67_{-0.92}^{+1.75}$ & $518/454$ & $0.03$ &  $2.57_{-0.03}^{+0.03}$ & $0.82_{-0.04}^{+0.04}$ & $7.20_{-0.05}^{+0.05}$ & $1.56\pm0.01$ \\

			& 0742590301$^{\dagger}$ & 56843.97 & $0.29_{-0.01}^{+0.01}$ & $0.24_{-0.01}^{+0.01}$ & $0.75_{-0.01}^{+0.03}$ & $2.53_{-0.02}^{+0.02}$ & $60.64_{-2.55}^{+3.18}$ & $572/433$ & $0.01$ &  $3.41_{-0.05}^{+0.05}$ & $3.59_{-0.03}^{+0.03}$ & $11.09_{-0.05}^{+0.05}$ & $2.40\pm0.01$ \\

			& 0742490101$^{\dagger}$ & 57111.19 & $0.27_{-0.01}^{+0.01}$ & $0.25_{-0.03}^{+0.02}$ & $1.09_{-0.05}^{+0.06}$ & $1.38_{-0.11}^{+0.13}$ & $53.63_{-2.82}^{+4.89}$ &  $360/266$ &  $0.01$ &  $1.81_{-0.03}^{+0.03}$ & $1.92_{-0.03}^{+0.03}$ & $5.91_{-0.05}^{+0.05}$ & $1.28\pm0.01$ \\		
			
			& 0803990101$^{\dagger}$ & 57918.89 & $0.29_{-0.01}^{+0.01}$ & $0.26_{-0.01}^{+0.01}$ & $1.07_{-0.02}^{+0.03}$ & $1.18_{-0.05}^{+0.06}$ & $33.82_{-2.54}^{+3.80}$ &  $682/486$ & $0.01$ &  $3.35_{-0.02}^{+0.02}$ & $4.95_{-0.03}^{+0.03}$ & $10.87_{-0.05}^{+0.05}$ & $2.35\pm0.01$ \\
			
			& 0803990301$^{\dagger}$ & 57996.68 & $0.29_{-0.02}^{+0.01}$ & $0.23_{-0.01}^{+0.01}$ & $1.30_{-0.02}^{+0.09}$ & $1.35_{-0.07}^{+0.08}$ & $44.47_{-3.12}^{+4.33}$ &  $478/422$ & $0.03$ &  $2.46_{-0.04}^{+0.04}$ & $2.48_{-0.04}^{+0.04}$ & $7.18_{-0.08}^{+0.09}$ & $1.55\pm0.02$ \\
			
			& 0803990601$^{\dagger}$ & 58096.46 & $0.27_{-0.01}^{+0.01}$ & $0.29_{-0.01}^{+0.01}$ & $1.09_{-0.03}^{+0.01}$ & $1.09_{-0.02}^{+0.05}$ & $34.52_{-0.70}^{+0.73}$ & $614/455$ & $0.01$ &  $3.14_{-0.04}^{+0.04}$ & $5.39_{-0.06}^{+0.06}$ & $9.86_{-0.08}^{+0.08}$ & $2.13\pm0.02$ \\
			
			\hline
			
			
			NGC 5408 X-1 & 0112290501$^{\dagger}$ & 52121.54 & $0.11^{*}$ & $0.18_{-0.01}^{+0.01}$ & $0.82_{-0.08}^{+0.09}$ & $-$ & $-$ & $128/130$ & $0.53$ &  $3.24_{-0.13}^{+0.13}$ & $0.86_{-0.05}^{+0.05}$ & $5.28_{-0.24}^{+0.24}$ & $1.46\pm0.06$ \\

			(d $=$ 4.8 ${\rm Mpc}^{e}$) & 0112291001$^{\dagger}$ & 52484.35 & $0.12_{-0.03}^{+0.04}$ & $0.19_{-0.02}^{+0.02}$ & $0.88_{-0.19}^{+0.42}$ & $-$ & $-$ & $125/136$ & $0.74$ &  $2.12_{-0.02}^{+0.02}$ & $0.89_{-0.06}^{+0.06}$ & $4.27_{-0.12}^{+0.12}$ & $1.18\pm0.03$ \\

			& 0112291201$^{\dagger}$ & 52667.01 & $0.13_{-0.05}^{+0.06}$ & $0.19_{-0.03}^{+0.03}$ & $1.08_{-0.15}^{+0.22}$ & $-$ & $-$ & $92/97$ & $0.62$ &  $2.01_{-0.11}^{+0.11}$ & $0.77_{-0.05}^{+0.05}$ & $3.11_{-0.15}^{+0.15}$ & $0.86\pm0.04$ \\

			& 0302900101 & 53748.77 & $0.11_{-0.01}^{+0.01}$ & $0.16_{-0.01}^{+0.01}$ & $0.76_{-0.04}^{+0.06}$ & $2.35_{-0.16}^{+0.14}$ & $7.93_{-2.27}^{+2.51}$ & $423/337$ & $0.01$ &  $2.21_{-0.02}^{+0.02}$ & $0.39_{-0.01}^{+0.01}$ & $4.99_{-0.03}^{+0.03}$ & $1.38\pm0.01$ \\
			
			& 0500750101 & 54478.79 & $0.11_{-0.01}^{+0.01}$ & $0.17_{-0.01}^{+0.01}$ & $0.77_{-0.04}^{+0.06}$ & $2.14_{-0.29}^{+0.25}$ & $6.61_{-2.96}^{+4.10}$ & $320/294$ & $0.14$ &  $1.84_{-0.03}^{+0.03}$ & $0.49_{-0.02}^{+0.02}$ & $4.29_{-0.05}^{+0.05}$ & $1.18\pm0.01$ \\
			
			& 0653380201 & 55394.13 & $0.11_{-0.01}^{+0.01}$ & $0.17_{-0.01}^{+0.01}$ & $0.89_{-0.09}^{+0.13}$ & $2.46_{-0.17}^{+0.15}$ & $14.08_{-3.84}^{+3.63}$ & $321/301$ & $0.21$ &  $2.26_{-0.02}^{+0.02}$ & $0.39_{-0.02}^{+0.02}$ & $5.34_{-0.04}^{+0.04}$ & $1.47\pm0.01$ \\
			
			& 0653380301 & 55396.13 & $0.09_{-0.01}^{+0.01}$ & $0.18_{-0.01}^{+0.01}$ & $0.99_{-0.12}^{+0.18}$ & $2.53_{-0.12}^{+0.13}$ & $19.75_{-3.61}^{+3.79}$ & $429/364$ & $0.01$ &  $1.87_{-0.01}^{+0.01}$ & $0.31_{-0.01}^{+0.01}$ & $4.48_{-0.03}^{+0.03}$ & $1.24\pm0.01$  \\
			
			& 0653380401 & 55587.67 & $0.10_{-0.01}^{+0.01}$ & $0.18_{-0.01}^{+0.01}$ & $0.81_{-0.04}^{+0.06}$ & $2.23_{-0.16}^{+0.19}$ & $8.73_{-2.50}^{+3.96}$ & $398/356$ & $0.06$ &  $1.97_{-0.02}^{+0.02}$ & $0.51_{-0.01}^{+0.01}$ & $4.49_{-0.03}^{+0.03}$ & $1.24\pm0.01$ \\
			
			& 0653380501 & 55589.65 & $0.12_{-0.01}^{+0.01}$& $0.18_{-0.01}^{+0.01}$ & $0.81_{-0.03}^{+0.03}$ & $1.30_{-0.14}^{+0.24}$ & $16.40_{-1.20}^{+3.02}$ & $402/341$ & $0.01$ &  $1.90_{-0.03}^{+0.03}$ & $0.94_{-0.02}^{+0.02}$ & $4.24_{-0.04}^{+0.04}$ & $1.17\pm0.01$ \\
			
			& 0723130301 & 56699.02 & $0.11_{-0.01}^{+0.01}$ & $0.16_{-0.01}^{+0.01}$ & $0.81_{-0.04}^{+0.05}$ & $1.45_{-0.09}^{+0.11}$ & $1.99_{-0.53}^{+1.07}$ & $280/244$ & $0.06$ &  $1.91_{-0.04}^{+0.04}$ & $0.71_{-0.02}^{+0.02}$ & $4.20_{-0.06}^{+0.06}$ & $1.16\pm0.02$ \\
			
			& 0723130401 & 56701.02 & $0.11_{-0.01}^{+0.02}$ & $0.18_{-0.02}^{+0.01}$ & $0.93_{-0.09}^{+0.12}$ & $2.34_{-0.25}^{+0.25}$ & $11.09_{-4.33}^{+5.08}$ & $268/262$ & $0.38$ &  $1.91_{-0.03}^{+0.03}$ & $0.44_{-0.02}^{+0.02}$ & $4.44_{-0.05}^{+0.05}$ & $1.22\pm0.01$ \\
			
			\hline
			
			
			NGC 6946 X-1 & 0500730201 & 54406.91 & $0.31_{-0.04}^{+0.04}$ & $0.17_{-0.01}^{+0.02}$ & $1.13_{-0.10}^{+0.11}$ & $2.40_{-0.22}^{+0.24}$ & $16.49_{-2.45}^{+3.52}$ & $93/104$ & $0.77$ &  $1.32_{-0.04}^{+0.04}$ & $0.15_{-0.02}^{+0.02}$ & $3.08_{-0.07}^{+0.07}$ & $2.20\pm0.05$  \\
			
			(d $=$ 7.72 ${\rm Mpc}^{b}$) & 0500730101 & 54412.94 & $0.27_{-0.03}^{+0.03}$ & $0.20_{-0.01}^{+0.01}$ & $0.84_{-0.13}^{+0.13}$ & $1.50_{-0.23}^{+0.35}$ & $4.64_{-0.93}^{+2.00}$ & $195/181$ & $0.21$ &  $1.19_{-0.06}^{+0.06}$ & $0.31_{-0.03}^{+0.03}$ & $2.66_{-0.06}^{+0.07}$ & $1.90\pm0.04$ \\
			
			& 0691570101 & 56221.74& $0.31_{-0.02}^{+0.03}$ & $0.18_{-0.01}^{+0.01}$ & $1.12_{-0.23}^{+0.40}$ & $2.41_{-0.20}^{+0.24}$ & $14.07_{-3.77}^{+4.02}$ & $325/330$ & $0.56$ & $1.83_{-0.03}^{+0.03}$ & $0.18_{-0.02}^{+0.02}$ & $4.07_{-0.05}^{+0.05}$ & $2.90\pm0.04$ \\
			
			& 0870830101 & 59038.93 & $0.32_{-0.06}^{+0.06}$ & $0.21_{-0.02}^{+0.03}$ & $1.30_{-0.14}^{+0.17}$ & $-$ & $-$ & $142/135$ & $0.32$ & $1.89_{-0.08}^{+0.08}$ & $0.67_{-0.04}^{+0.05}$ & $3.48_{-0.16}^{+0.15}$ & $2.48\pm0.11$ \\
			
			& 0870830201 & 59198.32 & $0.37_{-0.08}^{+0.08}$ & $0.18_{-0.02}^{+0.03}$ & $1.07_{-0.11}^{+0.14}$ & $-$ & $-$ & $87/82$ & $0.31$ & $2.85_{-0.19}^{+0.20}$ & $0.69_{-0.05}^{+0.05}$ & $5.34_{-0.37}^{+0.37}$ & $3.81\pm0.26$ \\
			
			& 0870830301 & 59308.22 & $0.28_{-0.05}^{+0.05}$ & $0.22_{-0.02}^{+0.03}$ & $1.30_{-0.12}^{+0.14}$ & $-$ & $-$ & $122/134$ & $0.75$ &  $1.44_{-0.06}^{+0.07}$ & $0.62_{-0.03}^{+0.03}$ & $2.74_{-0.06}^{+0.06}$ & $1.95\pm0.04$ \\
			
			& 0870830401 & 59359.89 & $0.32_{-0.06}^{+0.07}$ & $0.19_{-0.02}^{+0.03}$ & $1.22_{-0.12}^{+0.14}$ & $-$ & $-$ & $88/100$ & $0.78$ & $1.62_{-0.07}^{+0.07}$ & $0.56_{-0.03}^{+0.03}$ & $3.07_{-0.07}^{+0.08}$ & $2.19\pm0.06$ \\
			
			\hline
			
			
			IC 342 X-1 & 0093640901$^{\dagger}$ & 51951.06 & $0.58_{-0.06}^{+0.08}$ & $0.42_{-0.06}^{+0.08}$ & $2.78_{-0.27}^{+0.38}$ & $-$ & $-$ & $136/138$ & $0.52$ &  $0.91_{-0.08}^{+0.08}$ & $2.41_{-0.11}^{+0.11}$ & $3.91_{-0.18}^{+0.18}$ & $0.61\pm0.03$ \\
			
			(d $=$ 3.61 ${\rm Mpc}^{d}$) & 0206890101$^{\dagger}$ & 53055.31 & $0.63_{-0.03}^{+0.03}$ & $0.54_{-0.06}^{+0.06}$ & $2.13_{-0.11}^{+0.13}$ & $-$ & $-$ & $284/304$ & $0.79$ & $2.35_{-0.11}^{+0.11}$ & $5.83_{-0.13}^{+0.13}$ & $9.02_{-0.22}^{+0.23}$ & $1.41\pm0.03$ \\
			
			& 0206890201$^{\dagger}$ & 53234.82 & $0.65_{-0.03}^{+0.04}$ & $0.43_{-0.03}^{+0.03}$ & $2.68_{-0.15}^{+0.17}$ & $-$ & $-$ & $290/276$ & $0.26$ & $1.49_{-0.03}^{+0.03}$ & $2.57_{-0.05}^{+0.05}$ & $4.72_{-0.05}^{+0.05}$ & $0.74\pm0.01$ \\
			
			& 0693850601 & 56150.83 & $0.74_{-0.04}^{+0.04}$ & $0.29_{-0.02}^{+0.02}$ & $1.74_{-0.16}^{+0.18}$ & $1.30_{-0.13}^{+0.12}$ & $13.43_{-2.47}^{+2.93}$ & $330/360$ & $0.86$ & $2.09_{-0.07}^{+0.07}$ & $1.48_{-0.07}^{+0.07}$ & $5.96_{-0.08}^{+0.08}$ & $0.93\pm0.01$ \\
			
			& 0693851301$^{\dagger}$ & 56156.84 & $0.59_{-0.02}^{+0.02}$ & $0.48_{-0.02}^{+0.03}$ & $2.44_{-0.08}^{+0.09}$ & $-$ & $-$ & $402/377$ & $0.17$ & $1.46_{-0.04}^{+0.04}$ & $3.05_{-0.06}^{+0.06}$ & $5.13_{-0.06}^{+0.06}$ & $0.80\pm0.01$ \\
			
			\hline
			\hline
			
		\end{tabular}%
	}
	\begin{list}{}{}
		\item $^{\dagger}$Non-detection of QPO.
		\item $^\ast$Frozen parameter value.
		\item $^\boxtimes$$P_{\rm null} < 0.1$ yielded due to excess residuals at $0.56$ keV and $1$ keV in some of the observations of NGC 1313 X$-$1 and NGC 5408 X$-$1.
		\item $^{e}$\cite{Luangtip-etal2021}, $^c$\cite{Tully-etal2016}, $^{b}$\cite{Anand-etal2018}, $^{d}$\cite{Middleton-etal2015}.
	\end{list}
\end{table*}

We initially model the energy spectra of {\it XMM-Newton} in $0.3-10$ keV energy range using the phenomenological model combination \texttt{TBabs$\times$(diskbb $+$ diskbb)}. Here, \texttt{TBabs} \cite[]{Wilms-etal2000} accounts for the galactic absorption and two \texttt{diskbb} models \cite[]{Makishima-etal1986} correspond to the standard accretion disc components. We notice that the above model combinations is inadequate ($e.g.$, $\chi_{\rm red}^2 = 1.41$ for NGC 1313 X$-$1, ObsID-0693851201) in explaining the high energy tail of the spectra beyond $\sim 7$ keV. Hence, a \texttt{simpl} \cite[]{Steiner-etal2009} model that delineates the fractional scattering of the seed photons into power-law distribution is included in the spectral fitting. Such a model combination (hereafter M1) \texttt{TBabs$\times$(diskbb $+$ simpl$\times$diskbb)} is seen to describe the spectral shape of the BH-ULXs satisfactorily. For example, we obtain best spectral fit for NGC 1313 X$-$1 (ObsID-0693851201) with $\chi_{\rm red}^2=1.19$ that resulted relatively hotter disc temperature ($kT_{\rm in}^{\rm hard}$) $= 2.08_{-0.35}^{+0.46}$ keV, softer disc temperature ($kT_{\rm in}^{\rm soft}$) $= 0.23_{-0.02}^{+0.02}$ keV, photon index ($\Gamma_{\rm simpl}$) $=1.83_{-0.09}^{+0.16}$, seed photon scattered fraction ($f_{\rm scatt}$) $=20.67_{-0.92}^{+1.75}$ \%  and column density ($N_{\rm H}$) =$0.27_{-0.02}^{+0.02} \times 10^{22}$ atoms $\rm cm^{-2}$. Accordingly, we employ model M1 to fit the spectra of four BH-ULXs and obtain statistically acceptable fit. However, we mention that for few observations, \texttt{simpl} model is not required mostly due to the lack of good quality data at high energies (see Table \ref{table:disc_spec_parameters}). The model fitted parameters and estimated parameters are presented in Table \ref{table:disc_spec_parameters}.

Next, we examine the spectral behaviour of the four BH-ULXs considering physically motivated model. While doing so, we begin with the model combination \texttt{TBabs$\times$nthComp} to fit the spectra, where \texttt{nthComp} \cite[]{Zdziarski-etal1996,Zycki-etal1999} represents the thermally Comptonize continuum. The obtained fit yields poor $\chi_{\rm red}^2 = 2.87$ ($\chi^2/dof=1309/456$) for NGC 1313 X$-$1 (ObsID-0693851201) with noticeable residuals at lower energy. Hence, we include a \texttt{diskbb} model to fit the soft excess below $\sim 2$ keV. Further, we tie the seed photon temperature of \texttt{nthComp} to the inner disc temperature of \texttt{diskbb} \cite[]{Gladstone-etal2009}. Accordingly, the model combination \texttt{TBabs$\times$(nthComp $+$ diskbb)} (hereafter M2) provides a good fit ($\chi_{\rm red}^2=1.20$ for NGC 1313 X$-$1; ObsID-0693851201). In Fig. \ref{fig:residual_nth}, the variations of residuals ($\chi$) in units of $\sigma$ are presented in terms of energy in keV. In each panel of Fig. \ref{fig:residual_nth}, we choose one particular observation for a given source and obtain the variation of $\chi$ from the spectral fitting of that observation using model M2. Note that the source names as well as the observation details are mentioned in each panel. 

We further put efforts to fit the spectra with an alternative Comptonization model \texttt{compTT} \cite[]{Titarchuk-etal1994,Titarchuk-etal1995} that results statistically acceptable fits, however, the electron temperature remains unconstrained in most of the observations. Moreover, we adopt the slim disc prescription (\texttt{diskpbb} in \texttt{XSPEC} \cite[]{Mineshige-etal1994, Hirano-etal1995, Watarai-etal2000, Kubota-etal2004}) for spectral fitting which yields worse fit with $\chi_{\rm red}^2$ exceeding $2$ for all the sources. This evidently rules out the possibility of having slim accretion disc structure in these four BH-ULXs under consideration.

\subsection{Spectral Properties}

\begin{figure}
	\begin{center}
		\includegraphics[width=\columnwidth]{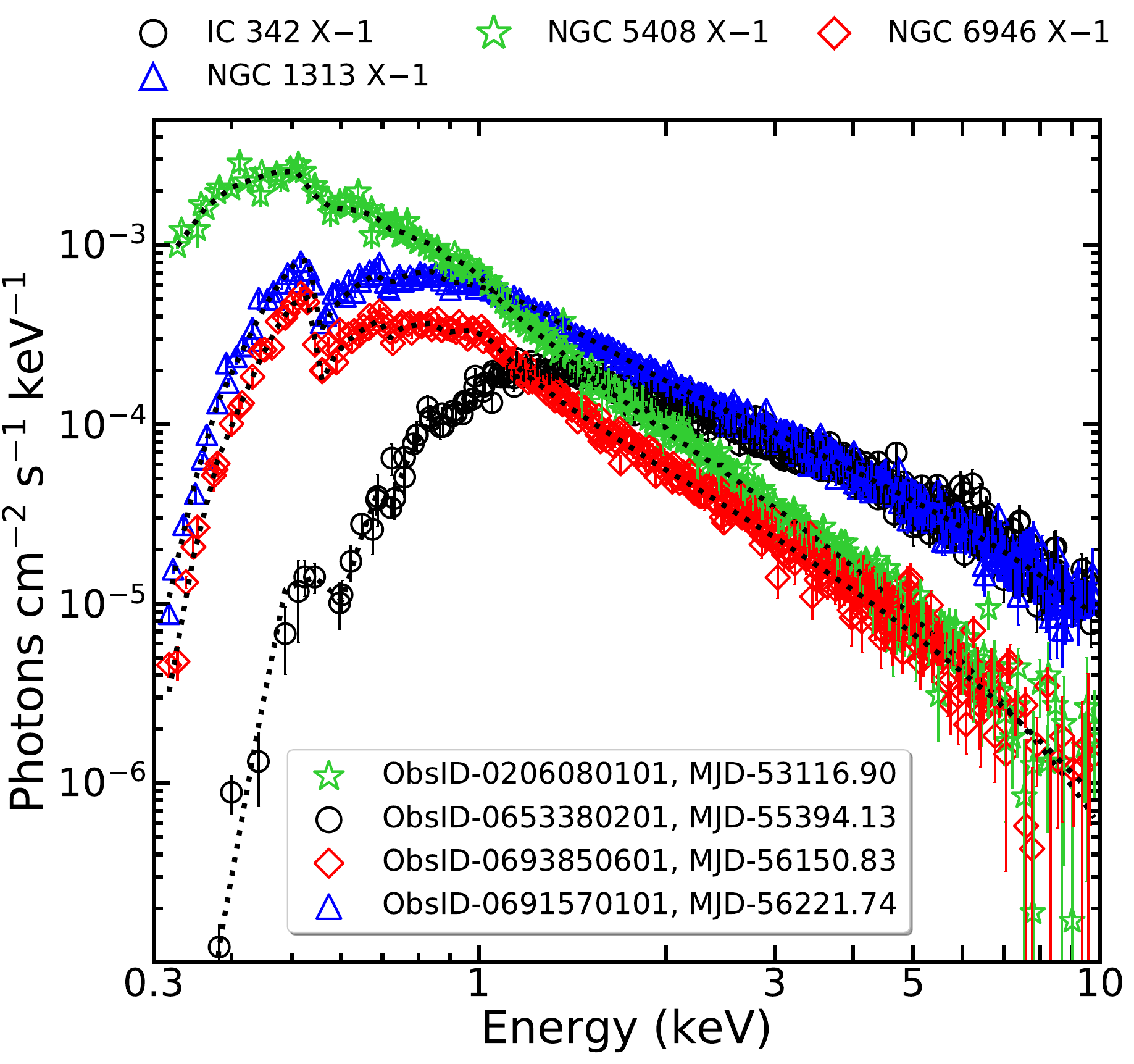}
	\end{center}
	\caption{Combined {\it EPIC-PN} and {\it EPIC-MOS} energy spectra ($0.3-10$ keV) of four BH-ULXs under consideration. Each spectrum is fitted with the model \texttt{TBabs}$\times$(\texttt{diskbb} + \texttt{nthComp}). The observation details are mentioned in the inset. See text for details.}
	\label{fig:spectra}
\end{figure}

In Fig. \ref{fig:spectra}, we display the energy spectra fitted with model combination M2  corresponding to the observations that exhibits maximum QPO frequency (see Table \ref{table:PDS_parameters}) for four BH-ULXs. The source names and observation details are marked in the figure. The best fitted model parameters and estimated parameters corresponding to the models M1 and M2 are tabulated in Table \ref{table:disc_spec_parameters} and \ref{table:Comp_spec_parameters}, respectively. We find that the model M1 satisfactorily describes all the spectra in terms of two disc temperatures varying in the range of $0.16_{-0.01}^{+0.01}-0.60_{-0.05}^{+0.09}$ keV and $0.75_{-0.01}^{+0.03}-3.13_{-0.21}^{+0.28}$ keV, respectively. In addition, the photon index ($\Gamma_{\rm simpl}$) and the scattered fraction of seed photons ($f_{\rm scatt}$) in \texttt{simpl} component of the model M1 are obtained in the range of $1.09_{-0.01}^{+0.15}-2.53_{-0.12}^{+0.13}$ and $1.99_{-0.53}^{+1.07}-60.64_{-2.55}^{+3.18}\%$, respectively. The spectral fitting using model M2 constrains the electron temperature ($kT_{\rm e}$) and \texttt{nthComp} photon index ($\Gamma_{\rm nth}$) as $1.62_{-0.12}^{+0.13}-3.76_{-0.67}^{+1.15}$ keV and $1.48_{-0.11}^{+0.04}-2.65_{-0.17}^{+0.19}$, respectively. The inner disc temperature ($kT_{\rm in}$) is found to be in the range of $0.14_{-0.02}^{+0.03}-0.54_{-0.06}^{+0.15}$ keV suggesting a cool accretion disc around the sources. Note that in few observations, we are unable to constrain the errors associated with the electron temperature and hence, it is kept fixed at its fitted values (see Table \ref{table:Comp_spec_parameters}). 

In order to understand the emission properties of BH-ULXs, we estimate the flux associated with each spectral component used in the model fitting. The flux is obtained by introducing the convolution model \texttt{cflux} (in \texttt{XSPEC}) after the absorption column density in the spectral fitting. While doing so, we freeze the normalizations of the spectral components at its best fitted values. With this, we compute the bolometric luminosity $L_{\rm bol}=F_{\rm bol}\times4\pi d^{2}$, where $F_{\rm bol}$ is the bolometric flux in $0.1-100$ keV energy range, and $d$ is the source distance. In Table \ref{table:disc_spec_parameters} and Table \ref{table:Comp_spec_parameters}, we present the computed $L_{\rm bol}$ for all the observations. We also compute the $L_{\rm disk}$ by calculating the bolometric disk flux $F_{\rm disk}$ in $0.1-100$ keV range as $L_{\rm disk}=F_{\rm disk} \times 4 \pi d^2$. Following  \cite{Zdziarski-etal1996,Majumder-etal2022}, we calculate the optical depth ($\tau$) and Compton y-parameter (y-par) for all the observations and obtain their ranges as $7 < \tau < 20$ and $1.16 \lesssim$ y-par $\lesssim 6.24$, respectively (see Table \ref{table:Comp_spec_parameters}). These findings eventually indicate the existence of an optically thick and `cool' corona surrounding the BH-ULXs under consideration.

\subsection{Long-term Spectral Evolution}
	
	\begin{figure}
		\begin{center}
			\includegraphics[width=\columnwidth]{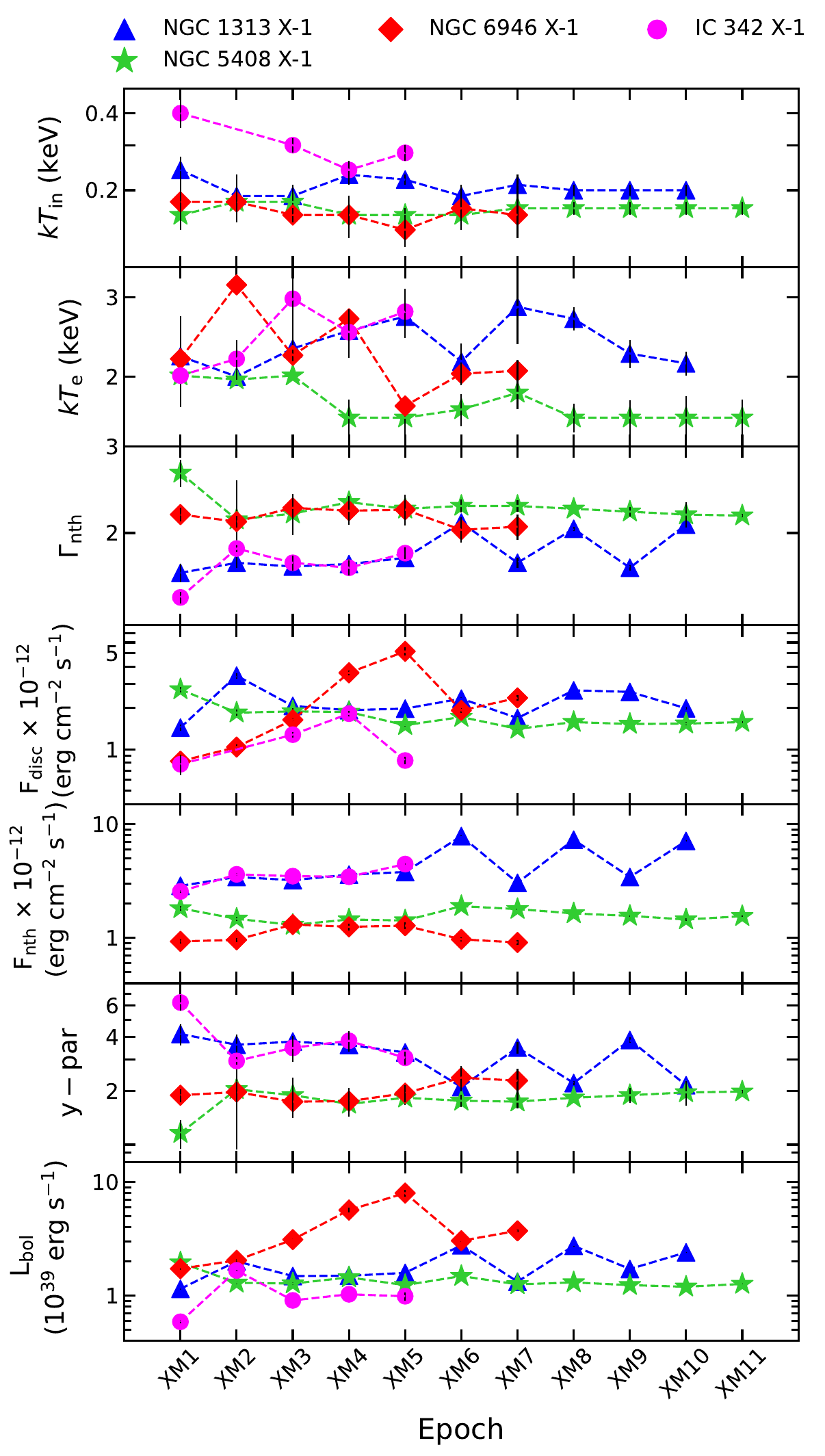}
		\end{center}
		\caption{Long-term evolution of the spectral parameters obtained from the best fitted energy spectra of all the sources using model M2. From top to bottom panels, the variation of disc temperature ($kT_{\rm in}$), electron temperature ($kT_e$), photon index ($\Gamma_{\rm nth}$), disc flux ($F_{\rm disc}$), Comptonization flux ($F_{\rm nth}$), Compton y-parameter (y-par) and bolometric luminosity ($L_{\rm bol}$) are depicted at different epochs. Filled points, namely triangle (blue), diamond (red), circle (magneta), and asterisk (green) joined with dashed lines denote results for NGC 1313 X$-$1, NGC 6946 X$-$1, IC 342 X$-$1, and NGC 5408 X$-$1, respectively.  See text for details.}
		\label{fig:spec_var}
	\end{figure}
		
Here, we investigate the long-term evolution of the spectral parameters of each source under consideration. While doing so, we make use of physically motivated model M2 for spectral modelling. In Fig. \ref{fig:spec_var}, we show the variation of inner disc temperature ($kT_{\rm in}$), electron temperature ($kT_{\rm e}$), photon index ($\Gamma_{\rm nth}$), disc flux ($F_{\rm disc}$), Comptonized flux ($F_{\rm nth}$), Compton y-parameter (y-par) and bolometric luminosity ($L_{\rm bol}$) over the epochs of observation from top to bottom panel, respectively. We observe that all the sources except NGC $5408$ X$-1$ show significant spectral evolution throughout the long-term monitoring for more than a decade. However, for NGC $5408$ X$-1$, marginal variation in spectral properties is observed. We find that NGC $1313$ X$-1$ and IC $342$ X$-1$ are dominated by the Comptonized flux ($F_{\rm nth} \sim 3-45 \times 10^{-12}$ erg cm$^{-2}$ s$^{-1}$) whereas NGC $6946$ X$-1$ and NGC $5408$ X$-1$ are seen to be disc dominated ($F_{\rm disc} \sim 1-5 \times 10^{-12}$ erg cm$^{-2}$ s$^{-1}$). In addition, for NGC $5408$ X$-1$ and NGC $6946$ X$-1$, the photon index is found to be more steeper ($\Gamma_{\rm nth} \gtrsim 2$) in comparison to other sources ($\Gamma_{\rm nth}\lesssim2$), which further confirms the softer nature of these two sources. We calculate the y-parameter and find that the amount of Comptonization is high (y-par $\sim2-6$) for NGC $1313$ X$-1$ and IC $342$ X$-1$ as compared to the remaining two sources. The electron temperature manifests an overall variation in the range of $\sim 2-4$ keV for all the sources. Interestingly, we find that the disc temperature demonstrates a minimal variation in all the sources except IC $342$ X$-1$, which possibly harbour variable disc of higher temperatures. Further, we observe that NGC $6946$ X$-1$ is the most luminous ULX in our sample, showing maximum long-term variation in the luminosity as well.

\begin{landscape}
	\begin{table}
		\caption{Best fitted spectral parameters obtained with the model \texttt{TBabs$\times$(diskbb $+$ nthComp)} for four ULXs in the sample. In the table, $n_{\rm H}$ is the hydrogen column density, $kT_{e}$ is the electron temperature in keV, $\Gamma_{\rm nth}$ is the \texttt{nthComp} photon index, $\tau$ is the optical depth and $kT_{\rm in}$ is the inner disc temperature in keV. $\chi^2/d.o.f$ and $P_{\rm null}$ refer the reduced $\chi^2$ and null hypothesis probability values. $F_{\rm disc}$ and $F_{\rm nth}$ are the flux values in units of erg cm$^{-2}$ s$^{-1}$ associated with \texttt{diskbb} and \texttt{nthComp} components, respectively. $F_{\rm bol}$, $L_{\rm bol}$ and $L_{\rm disc}$ are the bolometric flux, total luminosity and disc luminosity, respectively. All the errors are computed with $90$ per cent confidence level. See text for details.}
		\renewcommand{\arraystretch}{1.8}
		
		\label{table:Comp_spec_parameters}
		
		\resizebox{1.0\columnwidth}{!}{%
			\begin{tabular}{l @{\hspace{0.3cm}} l @{\hspace{0.2cm}} c @{\hspace{0.2cm}} c @{\hspace{0.2cm}} c @{\hspace{0.2cm}} c @{\hspace{0.22cm}} c @{\hspace{0.4cm}} c @{\hspace{0.4cm}} c @{\hspace{0.5cm}} c @{\hspace{0.5cm}} c @{\hspace{0.5cm}} c @{\hspace{0.3cm}} c @{\hspace{0.45cm}} c @{\hspace{0.35cm}} c @{\hspace{0.5cm}} c @{\hspace{0.6cm}} c @{\hspace{0.001cm}} c @{\hspace{0.001cm}} c }
				\hline
				\hline
				& & & & \multicolumn{2}{|c|}{Model fitted parameters} & & & \multicolumn{9}{|c|}{Estimated parameters} \\

				\cline{4-7}
				\cline{10-18}
				& & & & & & & & & & & \\
				Source & ObsID & MJD & $n_{\rm H}$ & $kT_{e}$ & $\Gamma_{\rm nth}$ & $kT_{in}$  & $\chi^2/d.o.f$ & $P_{null}$ $^{\boxtimes}$ & $F_{\rm disc}$ & $F_{\rm nth}$ & $F_{\rm bol}$ & $F_{\rm disc}$ & $L_{\rm bol}$ & $L_{\rm disc}$ & $\tau$ & y-par  \\
				
				& & & & & & & & & \multicolumn{2}{|c|}{($0.3-10$ keV)} & \multicolumn{2}{|c|}{($0.1-100$ keV)} & & & \\
				
				(Distance) & & & ($10^{22}$ ${\rm cm}^{-2}$) & (keV) & & (keV) & & & &  \multicolumn{2}{|c|}{($\tiny{10^{-12}}$ erg ${\rm cm}^{-2}$ ${\rm s}^{-1}$)} & & \multicolumn{2}{|c|}{($10^{40}$ erg ${\rm s}^{-1}$)} &  &  &  \\

				\hline
				
				
				NGC 1313 X-1 & 0106860101$^{\dagger}$ & 51834.14 & $0.25_{-0.04}^{+0.03}$ & $2.22_{-0.31}^{+0.51}$ & $1.66_{-0.09}^{+0.07}$ & $0.24_{-0.03}^{+0.04}$ & $128/130$ & $0.52$ &  $1.44_{-0.08}^{+0.08}$ & $2.87_{-0.09}^{+0.09}$ & $5.32_{-0.10}^{+0.11}$ & $2.00_{-0.11}^{+0.11}$ & $1.15\pm0.02$ & $0.43\pm0.02$ & $15.49\pm2.36$ & $4.16\pm0.57$  \\

				(d $=$ 4.25 ${\rm Mpc}$) & 0205230601$^{\dagger}$ & 53408.48 & $0.36_{-0.07}^{+0.08}$ & $2^{*}$ & $1.74_{-0.04}^{+0.08}$ & $0.19_{-0.03}^{+0.04}$ & $232/207$ & $0.11$ &  $3.44_{-0.16}^{+0.16}$ & $3.43_{-0.11}^{+0.11}$ & $9.31_{-0.24}^{+0.24}$ & $5.44_{-0.25}^{+0.25}$ & $2.01\pm0.05$ & $1.18\pm0.05$ & $15.22\pm1.07$ & $3.62\pm0.51$ \\

				& 0405090101 & 54023.98 & $0.31_{-0.01}^{+0.01}$ & $2.31_{-0.12}^{+0.14}$ & $1.71_{-0.03}^{+0.03}$ & $0.19_{-0.01}^{+0.01}$ & $532/446$ & $0.02$ & $2.07_{-0.02}^{+0.02}$ & $3.22_{-0.01}^{+0.02}$ & $6.90_{-0.05}^{+0.05}$ &  $3.21_{-0.05}^{+0.05}$ & $1.49\pm0.01$ & $0.69\pm0.01$ & $14.46\pm0.62$ & $3.77\pm0.21$  \\

				& 0693850501 & 56277.67 & $0.28_{-0.02}^{+0.02}$ & $2.53_{-0.17}^{+0.22}$ & $1.73_{-0.04}^{+0.04}$ & $0.23_{-0.01}^{+0.02}$ & $479/440$ & $0.10$ & $1.93_{-0.04}^{+0.04}$ & $3.60_{-0.05}^{+0.05}$ & $6.95_{-0.07}^{+0.07}$ &  $2.77_{-0.05}^{+0.05}$ & $1.50\pm0.02$ & $0.59\pm0.01$ & $13.51\pm0.81$ & $3.61\pm0.26$  \\

				& 0693851201 & 56283.66& $0.28_{-0.02}^{+0.02}$ & $2.72_{-0.19}^{+0.25}$ & $1.78_{-0.03}^{+0.03}$ & $0.22_{-0.01}^{+0.01}$ & $547/455$ & $0.01$ & $1.98_{-0.03}^{+0.03}$ & $3.80_{-0.04}^{+0.04}$ & $7.35_{-0.05}^{+0.05}$ &  $2.90_{-0.05}^{+0.05}$ & $1.59\pm0.01$ & $0.63\pm0.01$ & $12.42\pm0.71$ & $3.28\pm0.17$  \\

				& 0742590301$^{\dagger}$ & 56843.97 & $0.33_{-0.03}^{+0.03}$ & $2.16_{-0.16}^{+0.21}$ & $2.10_{-0.04}^{+0.04}$ & $0.19_{-0.01}^{+0.02}$ & $592/434$ & $0.01$ &  $2.34_{-0.06}^{+0.06}$ & $7.87_{-0.05}^{+0.05}$ & $12.86_{-0.06}^{+0.06}$ & $3.69_{-0.10}^{+0.10}$ & $2.78\pm0.01$ & $0.80\pm0.02$ & $11.14\pm0.67$ & $2.10\pm0.11$  \\

				& 0742490101$^{\dagger}$ & 57111.19 & $0.29_{-0.02}^{+0.03}$ & $2.86_{-0.50}^{+1.25}$ & $1.74_{-0.05}^{+0.04}$ & $0.21_{-0.02}^{+0.02}$ &  $355/267$ & $0.03$ &  $1.69_{-0.04}^{+0.04}$ & $3.05_{-0.04}^{+0.04}$ & $6.16_{-0.06}^{+0.06}$ & $2.51_{-0.06}^{+0.06}$ & $1.33\pm0.01$ & $0.54\pm0.01$ & $12.51\pm3.06$ & $3.49\pm0.26$  \\

				& 0803990101$^{\dagger}$ & 57918.89 & $0.34_{-0.01}^{+0.01}$ & $2.69_{-0.14}^{+0.16}$ & $2.04_{-0.02}^{+0.02}$ & $0.20_{-0.01}^{+0.01}$ & $684/540$ & $0.01$ &  $2.69_{-0.04}^{+0.04}$ & $7.31_{-0.03}^{+0.03}$ & $12.74_{-0.05}^{+0.05}$ &  $4.16_{-0.07}^{+0.07}$ & $2.75\pm0.01$ & $0.90\pm0.02$ & $10.25\pm0.37$ & $2.21\pm0.06$  \\

				& 0803990301$^{\dagger}$ & 57996.68 & $0.33_{-0.02}^{+0.02}$ & $2.25_{-0.13}^{+0.16}$ & $1.70_{-0.02}^{+0.02}$ & $0.20_{-0.01}^{+0.01}$ & $471/420$ & $0.05$ &  $2.62_{-0.05}^{+0.05}$ & $3.43_{-0.01}^{+0.01}$ & $7.99_{-0.09}^{+0.09}$ &  $4.07_{-0.07}^{+0.07}$ & $1.72\pm0.02$ & $0.88\pm0.02$ & $14.80\pm0.64$ & $3.85\pm0.15$  \\

				& 0803990601$^{\dagger}$ & 58096.46 & $0.32_{-0.01}^{+0.02}$ & $2.14_{-0.11}^{+0.13}$ & $2.08_{-0.02}^{+0.02}$ & $0.20_{-0.01}^{+0.01}$ & $642/456$ & $0.01$ &  $1.99_{-0.05}^{+0.05}$  & $7.13_{-0.07}^{+0.07}$ & $11.16_{-0.09}^{+0.09}$ & $2.97_{-0.08}^{+0.08}$ & $2.41\pm0.02$ & $0.64\pm0.02$ & $11.35\pm0.41$ & $2.15\pm0.06$  \\
				
				\hline

				
				NGC 5408 X-1 & 0112290501$^{\dagger}$ & 52121.54 & $0.11^{*}$ & $2.01^{*}$ & $2.65_{-0.17}^{+0.19}$ & $0.16_{-0.01}^{+0.01}$ & $125/129$ & $0.57$ & $2.73_{-0.12}^{+0.12}$ & $1.82_{-0.09}^{+0.09}$ & $7.11_{-0.20}^{+0.20}$ & $4.78_{-0.21}^{+0.21}$ & $1.96\pm0.06$ & $1.32\pm0.06$ & $8.62\pm0.77$  & $1.16\pm0.21$  \\

				(d $=$ 4.8 ${\rm Mpc}$) & 0112291001$^{\dagger}$ & 52484.35 & $0.12_{-0.03}^{+0.04}$ & $1.97^{*}$ & $2.13_{-0.45}^{+0.43}$ & $0.18_{-0.03}^{+0.03}$ & $126/136$ & $0.71$ & $1.85_{-0.09}^{+0.09}$ & $1.47_{-0.07}^{+0.07}$ & $4.72_{-0.13}^{+0.13}$ & $3.01_{-0.14}^{+0.14}$ & $1.30\pm0.04$ & $0.83\pm0.04$ & $11.51\pm3.26$ & $2.04\pm1.10$  \\

				& 0112291201$^{\dagger}$ & 52667.01 & $0.13_{-0.05}^{+0.06}$ & $2.01^{*}$ & $2.19_{-0.21}^{+0.21}$ & $0.18_{-0.03}^{+0.03}$ & $92/96$ & $0.58$ & $1.89_{-0.11}^{+0.11}$ & $1.30_{-0.09}^{+0.09}$ & $4.64_{-0.16}^{+0.16}$ & $3.10_{-0.17}^{+0.17}$ & $1.28\pm0.04$ & $0.85\pm0.05$ & $10.97\pm1.39$ & $1.89\pm0.48$  \\

				& 0302900101 & 53748.77& $0.12_{-0.01}^{+0.01}$ & $1.62_{-0.14}^{+0.16}$ & $2.31_{-0.01}^{+0.07}$ & $0.16_{-0.01}^{+0.01}$ & $438/338$ & $0.01$ & $1.87_{-0.01}^{+0.01}$ & $1.45_{-0.01}^{+0.01}$ & $5.27_{-0.02}^{+0.02}$ & $3.36_{-0.04}^{+0.04}$ & $1.45\pm0.01$ & $0.93\pm0.01$  & $11.54\pm0.78$ & $1.69\pm0.13$  \\

				& 0500750101 & 54478.79 & $0.12_{-0.01}^{+0.01}$ & $1.62_{-0.14}^{+0.18}$ & $2.24_{-0.02}^{+0.08}$ & $0.16_{-0.01}^{+0.01}$ & $333/295$ & $0.07$ & $1.50_{-0.03}^{+0.03}$ & $1.42_{-0.03}^{+0.03}$ & $4.49_{-0.05}^{+0.05}$ & $2.65_{-0.04}^{+0.04}$ & $1.24\pm0.01$ & $0.73\pm0.01$  & $12.01\pm0.93$ & $1.83\pm0.17$  \\

				& 0653380201 & 55394.13 & $0.11_{-0.01}^{+0.01}$ & $1.69_{-0.14}^{+0.19}$ & $2.27_{-0.05}^{+0.07}$ & $0.16_{-0.01}^{+0.01}$ & $334/303$ & $0.11$ & $1.72_{-0.02}^{+0.02}$ & $1.90_{-0.02}^{+0.02}$ & $5.40_{-0.04}^{+0.04}$ & $2.98_{-0.07}^{+0.07}$ & $1.49\pm0.01$ & $0.82\pm0.02$  & $11.53\pm0.86$ & $1.76\pm0.14$  \\

				& 0653380301 & 55396.13 & $0.10_{-0.01}^{+0.01}$ & $1.84_{-0.15}^{+0.20}$ & $2.27_{-0.07}^{+0.07}$ & $0.17_{-0.01}^{+0.01}$ & $419/366$ & $0.03$ & $1.41_{-0.02}^{+0.02}$ & $1.79_{-0.02}^{+0.02}$ & $4.57_{-0.03}^{+0.03}$ & $2.34_{-0.05}^{+0.05}$ & $1.26\pm0.01$ & $0.65\pm0.01$ & $11.00\pm0.80$ & $1.74\pm0.14$  \\
				
				\hline
				\hline
				
			\end{tabular}%
		}
		
		\begin{list}{}{}
			\item $^{\dagger}$Non-detection of QPO.
			\item $^\ast$Frozen parameter value.
			\item $^\boxtimes$$P_{\rm null} < 0.1$ yielded due to excess residuals at $0.56$ keV and $1$ keV in some of the observations of NGC 1313 X$-$1 and NGC 5408 X$-$1.
		\end{list}
		
	\end{table}
\end{landscape}

\begin{landscape}
	\begin{table}
		
		\contcaption{}
		\renewcommand{\arraystretch}{1.8}
		
		\resizebox{1.0\columnwidth}{!}{%
			\begin{tabular}{l @{\hspace{0.3cm}} l @{\hspace{0.2cm}} c @{\hspace{0.2cm}} c @{\hspace{0.2cm}} c @{\hspace{0.2cm}} c @{\hspace{0.22cm}} c @{\hspace{0.4cm}} c @{\hspace{0.4cm}} c @{\hspace{0.5cm}} c @{\hspace{0.5cm}} c @{\hspace{0.5cm}} c @{\hspace{0.3cm}} c @{\hspace{0.45cm}} c @{\hspace{0.35cm}} c @{\hspace{0.5cm}} c @{\hspace{0.6cm}} c @{\hspace{0.001cm}} c @{\hspace{0.001cm}} c}
				\hline
				\hline
				& & & & \multicolumn{2}{|c|}{Model fitted parameters} & & & \multicolumn{9}{|c|}{Estimated parameters} \\

				\cline{4-7}
				\cline{10-18}
				& & & & & & & & & & & & \\
				Source & ObsID & MJD & $n_{\rm H}$ & $kT_{e}$ & $\Gamma_{\rm nth}$ & $kT_{in}$  & $\chi^2/d.o.f$ & $P_{null}$$^{\boxtimes}$ & $F_{\rm disc}$ & $F_{\rm nth}$ & $F_{\rm bol}$ & $F_{\rm disc}$ & $L_{\rm bol}$ & $L_{\rm disc}$ & $\tau$ & y-par &  \\
				
				
				& & & & & & & & & \multicolumn{2}{|c|}{($0.3-10$ keV)} & \multicolumn{2}{|c|}{($0.1-100$ keV)} & & \\
				
				(Distance) & & & ($10^{22}$ ${\rm cm}^{-2}$) & (keV) & & (keV) & & &  &  \multicolumn{2}{|c|}{($\tiny{10^{-12}}$ erg ${\rm cm}^{-2}$ ${\rm s}^{-1}$)} & & \multicolumn{2}{|c|}{($10^{40}$ erg ${\rm s}^{-1}$)} & & &  \\

				\hline
				
				& 0653380401 & 55587.67 & $0.11_{-0.01}^{+0.01}$ & $1.62_{-0.12}^{+0.13}$ & $2.24_{-0.01}^{+0.01}$ & $0.17_{-0.01}^{+0.01}$ & $404/358$ & $0.05$ & $1.58_{-0.02}^{+0.02}$ & $1.64_{-0.02}^{+0.02}$ & $4.75_{-0.03}^{+0.03}$ & $2.68_{-0.06}^{+0.06}$ & $1.31\pm0.02$ & $0.74\pm0.02$ & $12.01\pm0.54$ & $1.83\pm0.03$  \\

				& 0653380501 & 55589.65 & $0.11_{-0.01}^{+0.01}$& $1.62_{-0.15}^{+0.17}$ & $2.21_{-0.01}^{+0.08}$ & $0.17_{-0.01}^{+0.01}$ & $415/342$ & $0.01$ & $1.53_{-0.02}^{+0.02}$ & $1.56_{-0.02}^{+0.02}$ & $4.50_{-0.04}^{+0.04}$ & $2.56_{-0.03}^{+0.03}$ & $1.24\pm0.01$ & $0.71\pm0.01$ & $12.22\pm0.92$ & $1.89\pm0.18$  \\

				& 0723130301 & 56699.02 & $0.12_{-0.01}^{+0.01}$ & $1.62_{-0.19}^{+0.24}$ & $2.18_{-0.02}^{+0.13}$ & $0.17_{-0.01}^{+0.01}$ & $297/244$ & $0.02$ & $1.54_{-0.04}^{+0.04}$ & $1.45_{-0.04}^{+0.04}$ & $4.36_{-0.06}^{+0.06}$ & $2.59_{-0.05}^{+0.05}$ & $1.20\pm0.02$ & $0.71\pm0.01$ & $12.44\pm1.41$ & $1.96\pm0.31$  \\

				& 0723130401 & 56701.02 & $0.11_{-0.01}^{+0.01}$ & $1.62_{-0.16}^{+0.18}$& $2.17_{-0.02}^{+0.02}$ & $0.17_{-0.01}^{+0.01}$ & $272/263$ & $0.34$ & $1.58_{-0.03}^{+0.03}$ & $1.55_{-0.03}^{+0.03}$ & $4.62_{-0.04}^{+0.04}$ & $2.67_{-0.05}^{+0.05}$ & $1.27\pm0.01$ & $0.74\pm0.01$ & $12.52\pm0.78$ & $1.98\pm0.05$  \\
				
				
				\hline
				
				NGC 6946 X-1 & 0500730201 & 54406.91 & $0.28_{-0.10}^{+0.11}$ & $2.19^{*}$ & $2.18_{-0.07}^{+0.06}$ & $0.18_{-0.02}^{+0.04}$ & $179/203$ & $0.88$ & $0.82_{-0.04}^{+0.04}$ & $0.93_{-0.04}^{+0.04}$ & $2.42_{-0.06}^{+0.06}$ & $1.29_{-0.06}^{+0.06}$ & $1.73\pm0.04$ & $0.93\pm0.04$ & $10.52\pm0.45$ & $1.89\pm0.16$  \\			
				
				(d $=$ 7.72 ${\rm Mpc}$) & 0500730101 & 54412.94 & $0.28_{-0.05}^{+0.07}$ & $3.20^{*}$ & $2.11_{-0.06}^{+0.11}$ & $0.18_{-0.03}^{+0.03}$ & $163/180$ & $0.81$ &  $1.04_{-0.05}^{+0.05}$ & $0.96_{-0.04}^{+0.04}$ & $2.87_{-0.07}^{+0.07}$ & $1.69_{-0.08}^{+0.08}$ & $2.05\pm0.05$ & $1.21\pm0.06$ & $8.87\pm0.64$ & $1.97\pm0.28$  \\
				
				& 0691570101 & 56221.74 & $0.32_{-0.05}^{+0.04}$ & $2.23_{-0.07}^{+0.07}$ & $2.25_{-0.15}^{+0.14}$ & $0.16_{-0.01}^{+0.01}$ & $326/331$ & $0.57$ & $1.64_{-0.04}^{+0.04}$ & $1.31_{-0.03}^{+0.03}$ & $4.38_{-0.10}^{+0.10}$ & $2.77_{-0.06}^{+0.06}$ & $3.12\pm0.07$ & $1.98\pm0.04$ & $9.98\pm0.84$ & $1.74\pm0.30$  \\

				& 0870830101 & 59038.93 & $0.43_{-0.09}^{+0.10}$ & $2.69^{*}$ & $2.22_{-0.14}^{+0.15}$ & $0.16_{-0.02}^{+0.03}$ & $135/135$ & $0.48$ & $3.61_{-0.16}^{+0.17}$ & $1.25_{-0.08}^{+0.09}$ & $7.96_{-0.36}^{+0.36}$ & $6.47_{-0.29}^{+0.31}$ & $5.68\pm0.26$ & $4.61\pm0.22$ & $9.14\pm0.83$ & $1.75\pm0.32$ &  \\
				
				& 0870830201 & 59198.32 & $0.47_{-0.11}^{+0.12}$ & $1.72^{*}$ & $2.23_{-0.16}^{+0.18}$ & $0.14_{-0.02}^{+0.03}$ & $84/82$ & $0.42$ & $5.17_{-0.36}^{+0.36}$ & $1.28_{-0.09}^{+0.09}$ & $11.23_{-0.81}^{+0.82}$ & $10.17_{-0.71}^{+0.74}$ & $8.01\pm0.58$ & $7.25\pm0.51$ & $12.37\pm1.18$ & $1.94\pm0.22$ &  \\
				
				& 0870830301 & 59308.22 & $0.34_{-0.07}^{+0.08}$ & $1.62^{*}$ & $2.03_{-0.12}^{+0.12}$ & $0.17_{-0.02}^{+0.03}$ & $119/134$ & $0.81$ &  $1.92_{-0.08}^{+0.08}$ & $0.97_{-0.04}^{+0.05}$ & $4.27_{-0.09}^{+0.10}$ & $3.16_{-0.14}^{+0.15}$ & $3.05\pm0.07$ & $2.25\pm0.08$ & $13.69\pm1.10$ & $2.37\pm0.38$  & \\
				
				& 0870830401 & 59359.89 & $0.37_{-0.08}^{+0.10}$ & $1.62^{*}$ & $2.06_{-0.12}^{+0.13}$ & $0.16_{-0.02}^{+0.03}$ & $85/100$ & $0.85$ & $2.37_{-0.10}^{+0.11}$ & $0.91_{-0.04}^{+0.04}$ & $5.23_{-0.12}^{+0.12}$ & $4.18_{-0.19}^{+0.20}$ & $3.73\pm0.09$ & $2.98\pm0.14$ & $13.42\pm1.14$ & $2.28\pm0.39$ & \\

				\hline
					
				
				IC 342 X-1 & 0093640901$^{\dagger}$ & 51951.06 & $0.58_{-0.08}^{+0.09}$ & $2.01^{*}$ & $1.48_{-0.11}^{+0.04}$ & $0.40_{-0.04}^{+0.05}$ & $136/138$ & $0.53$ &  $0.78_{-0.12}^{+0.13}$ & $2.56_{-0.08}^{+0.08}$ & $3.80_{-0.12}^{+0.12}$ & $0.89_{-0.04}^{+0.04}$ & $0.59\pm0.02$ & $0.14\pm0.01$ & $19.94\pm2.78$ & $6.24\pm0.63$  \\

				(d $=$ 3.61 ${\rm Mpc}$) & 0206890101$^{\dagger}$ & 53055.31 & $0.71_{-0.02}^{+0.02}$ & $2.19_{-0.16}^{+0.22}$ & $1.86_{-0.03}^{+0.03}$ & $-$ & $282/305$ & $0.82$ & $-$ & $3.62_{-0.09}^{+0.09}$ & $10.73_{-0.17}^{+0.17}$ & $-$ & $1.67\pm0.03$ & $-$ & $13.11\pm0.79$ & $2.94\pm0.14$  \\

				& 0206890201$^{\dagger}$ & 53234.82 & $0.74_{-0.07}^{+0.07}$ & $2.98_{-0.61}^{+1.56}$ & $1.74_{-0.09}^{+0.06}$ & $0.30_{-0.02}^{+0.02}$ & $287/275$ & $0.29$ &  $1.28_{-0.06}^{+0.06}$ & $3.49_{-0.06}^{+0.06}$ & $5.83_{-0.12}^{+0.12}$ & $1.68_{-0.05}^{+0.05}$ & $0.91\pm0.02$ & $0.26\pm0.01$ & $12.23\pm3.69$ & $3.48\pm0.57$  \\

				& 0693850601 & 56150.83 & $0.78_{-0.07}^{+0.08}$ & $2.51_{-0.23}^{+0.31}$ & $1.70_{-0.07}^{+0.06}$ & $0.24_{-0.01}^{+0.02}$ & $336/361$ & $0.82$ & $1.82_{-0.08}^{+0.08}$ & $3.45_{-0.06}^{+0.06}$ & $6.59_{-0.09}^{+0.09}$ &  $2.55_{-0.07}^{+0.07}$ & $1.03\pm0.01$ & $0.40\pm0.01$ & $13.94\pm1.31$ & $3.81\pm0.50$  \\

				& 0693851301$^{\dagger}$ & 56156.84 & $0.68_{-0.05}^{+0.06}$ & $2.79_{-0.35}^{+0.52}$ & $1.82_{-0.03}^{+0.05}$  & $0.28_{-0.02}^{+0.02}$ & $394/376$ & $0.25$ & $0.83_{-0.04}^{+0.05}$ & $4.46_{-0.04}^{+0.04}$ & $6.32_{-0.05}^{+0.05}$ & $1.11_{-0.06}^{+0.06}$ & $0.99\pm0.08$ & $0.17\pm0.01$ & $11.84\pm1.32$ & $3.06\pm0.26$  \\

				\hline
				\hline
				
			\end{tabular}%
		}
		\begin{list}{}{}
			\item $^\ast$Frozen parameter value.
			\item $^\boxtimes$$P_{\rm null} < 0.1$ yielded due to excess residuals at $0.56$ keV and $1$ keV in some of the observations of NGC 1313 X$-$1 and NGC 5408 X$-$1.
		\end{list}
		
	\end{table}
\end{landscape}

\section{Correlations among Spectro-temporal observables}

\label{s:spec-temp}

\begin{figure}
	\begin{center}
		\includegraphics[width=\columnwidth]{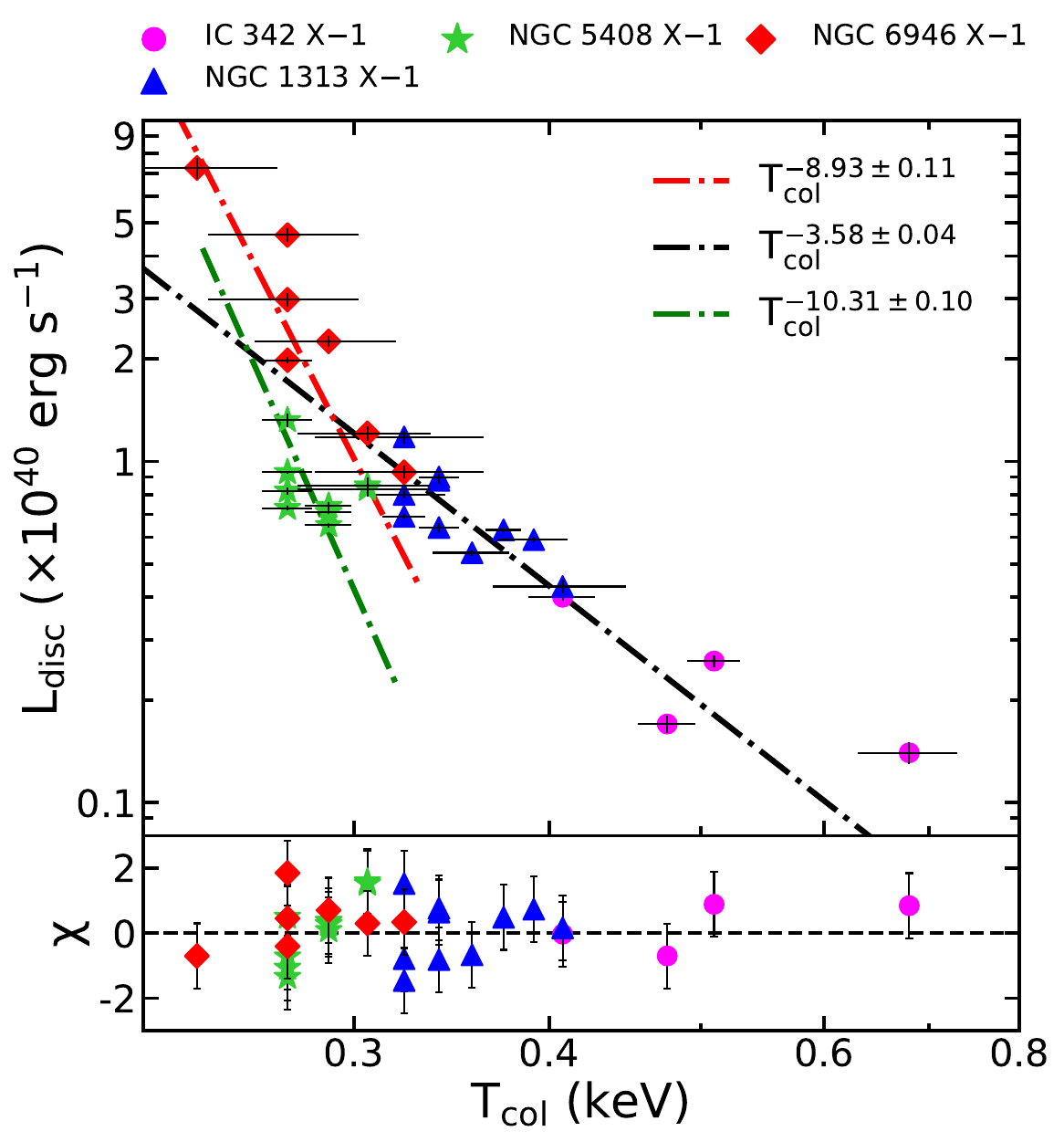}
	\end{center}
	\caption{Variation of bolometric disc luminosity ($L_{\rm disc}$) with color corrected inner disc temperature ($T_{\rm col}$) of four BH-ULXs. The results are obtained from the spectral fitting with the model M2. The filled points with different colors and styles denote various BH-ULXs as marked at the top of the figure. Dot-dashed lines represent best fitted function $L_{\rm disc} \propto T_{\rm col}^{\alpha}$. The blue shaded region indicates the $2 \sigma$ interval of the best fit. Bottom panel denotes the variation of residuals in units of $\sigma$. See text for details.
	}
	\label{fig:lum_temp}
\end{figure}

In this section, we examine the correlations among the various observables obtained from both timing and spectral analyses of the four BH-ULXs. With this, we put efforts to understand the spectro-temporal properties of these sources.

\subsection{Correlation in $L_{\rm disc}-T_{\rm col}$ Plane}

We carry out the correlation study of bolometric disc luminosity ($L_{\rm disc}$) and color corrected inner disc temperature ($T_{\rm col}$). Following \cite{Done-etal2008}, we compute $T_{\rm col} = f \times T_{\rm in}$ for spectral fitting with model M2 (see Table \ref{table:Comp_spec_parameters}), where $f~(=1.7)$ refers the spectral hardening factor. The obtained results are presented in Fig. \ref{fig:lum_temp}, where $L_{\rm disc}$ is plotted as function of $T_{\rm col}$. We calculate the Pearson correlation coefficient ($\rho$) to deduce the correlation between $L_{\rm disc}$ and $T_{\rm col}$ and observe an anti-correlation relation with $\rho \sim -0.81$ for NGC 1313 X$-$1 and IC 342 X$-$1, and $\rho \sim -0.89$ for NGC 6946 X$-$1. However, a weak negative correlation with $\rho \sim -0.27$ is observed for NGC 5408 $X-1$. Further, we run a MCMC simulation using the Goodman-Weare algorithm \cite[]{Goodman-etal2010} with a chain length of $20000$ and check for possible correlation (see Appendix-B) between disc flux ($F_{\rm disc}$) and inner disc temperature ($T_{\rm in}$). We find that $F_{\rm disc}$ depends on $T_{\rm in}$ for each source that rules out degeneracy and possibly causes the observed correlation in $L_{\rm disc}-T_{\rm col}$ plane.. Note that the above correlations are suitably described with an empirical power-law $L_{\rm disc} \propto T_{\rm col}^{\alpha}$ \cite[]{Rybicki-etal1979}, where $\alpha$ denotes the power-law index. A combined fitting for NGC 1313 X$-$1 and IC 342 X$-$1 yields $\alpha = -3.58 \pm 0.04$, whereas a more steeper index $\alpha = -8.93 \pm 0.11$ is obtained for NGC 6946 X$-$1. However, the fitting of the distribution in temperature-luminosity plane of NGC 5408 X-1 results $\alpha = -10.31 \pm 0.10$. The best fit power-law distributions are shown in upper panel of Fig. \ref{fig:lum_temp} along with the residuals in the lower panel.

\begin{figure}
	\begin{center}
		\includegraphics[width=\columnwidth]{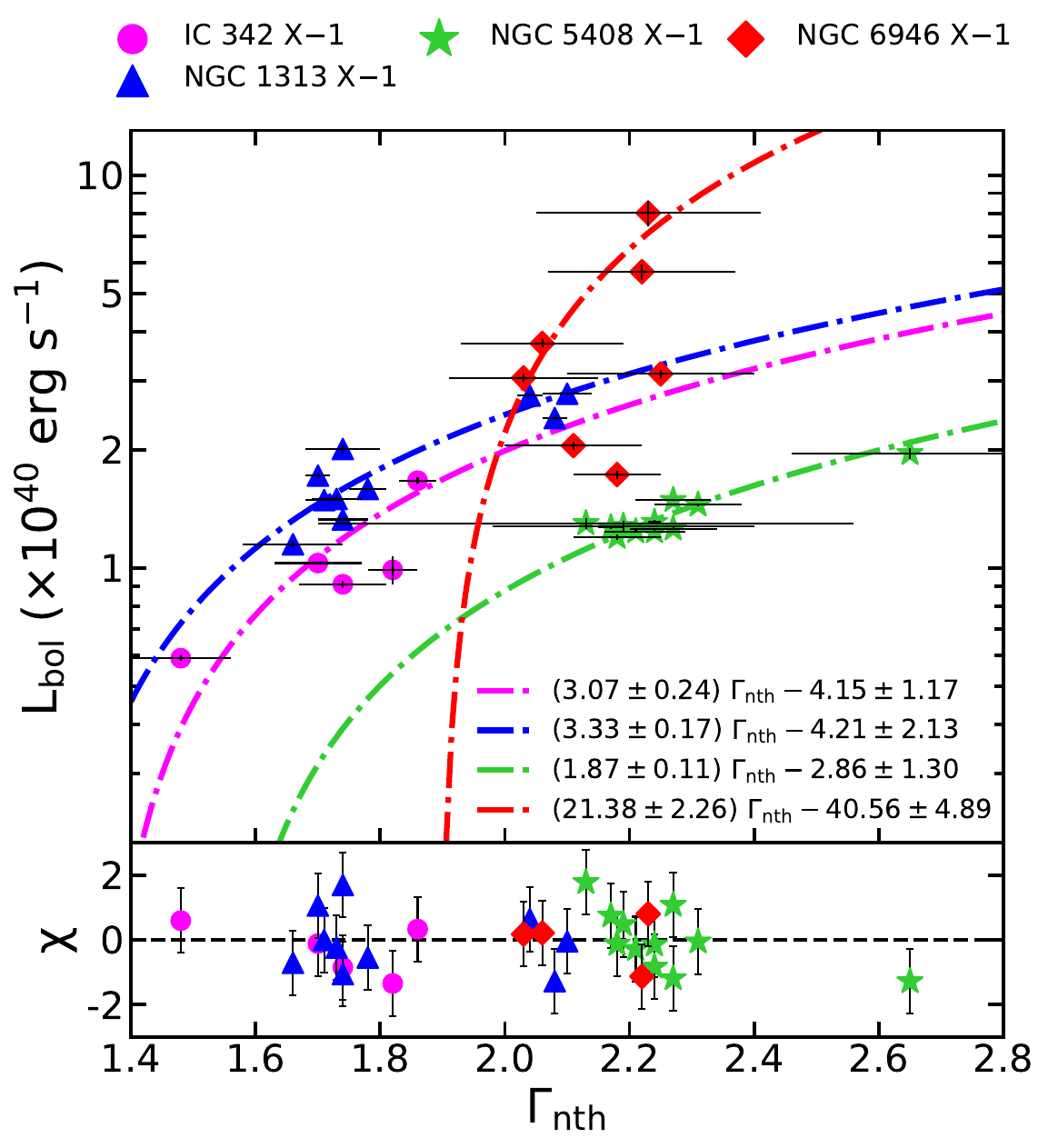}
	\end{center}
	\caption{Variation of total bolometric luminosity ($L_{\rm bol}$) with \texttt{nthComp} photon index ($\Gamma_{\rm nth}$) for four BH-ULXs. Filled points having different colors and styles are for different sources as marked at the top of the figure. Dot-dashed curves represent the best fitted linear functions for individual sources. Bottom panel denotes the variation of residuals in units of $\sigma$. See text for details.}
	\label{fig:gamma_lum}
\end{figure}

\subsection{Correlation in $L_{\rm bol}-\Gamma_{\rm nth}$ Plane}

We investigate the correlation between $L_{\rm bol}$ and the \texttt{nthComp} photon index ($\Gamma_{\rm nth}$) for four BH-ULXs. The obtained results are depicted in Fig. \ref{fig:gamma_lum}, where the variation of $L_{\rm bol}$ with $\Gamma_{\rm nth}$ is presented for each of the sources. Here, we observe strong positive correlation between $L_{\rm bol}$ and $\Gamma_{\rm nth}$ with Pearson correlation coefficient $+0.81$, $+0.92$, $+0.94$, and $+0.92$ for IC 342 X$-$1, NGC 1313 X$-$1, NGC 5408 X$-$1, and NGC 6946 X$-$1, respectively. Such a correlation is quantified using linear relation as $L_{\rm bol} = A \Gamma_{\rm nth} + C$, where $A$ and $C$ are source specific arbitrary constants. In the figure, the fitted functional relations are shown using dot-dashed curves in magenta, green, red, and blue for four BH-ULXs. Indeed, for NGC 6946 X$-$1, three data points at lower luminosities are excluded during fitting as they appears to be outliers. Note that the best-fitted points with higher luminosity belong to the observations of epoch XM5 to XM8 (See Table \ref{table: Obs_details}) during which the source shows a purely disc-dominated spectral state (see Fig. \ref{fig:spec_var}). The estimated values of these constants are obtained in the range of $1.87 < A < 21.38$ and $2.86 < C < 40.56$, respectively for all the sources as indicated in the Fig. \ref{fig:gamma_lum}.

\begin{figure}
	\begin{center}
		\includegraphics[width=\columnwidth]{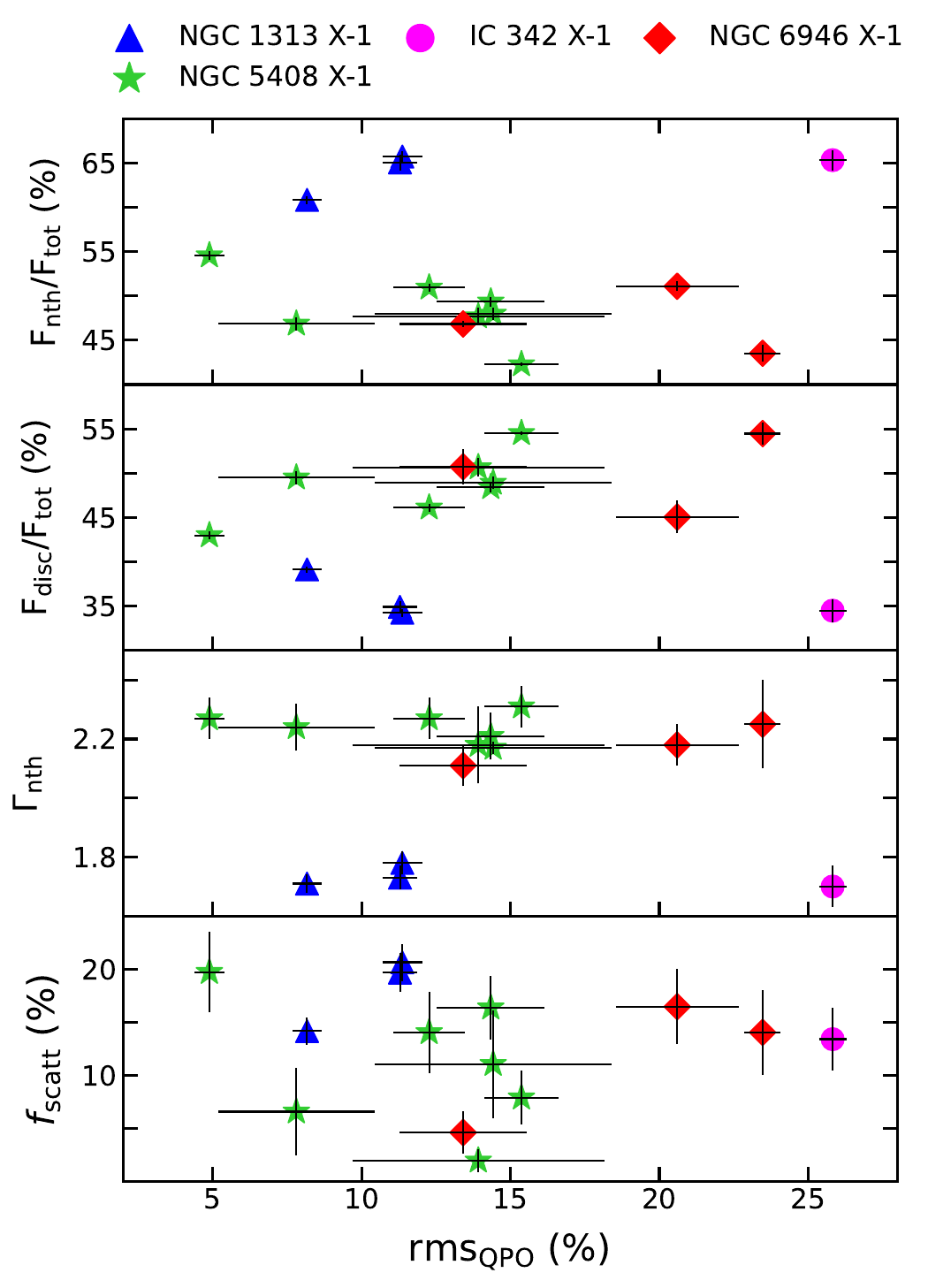}
	\end{center}
	\caption{Variations of the percentage of Comptonized flux ($F_{\rm nth}$) and disc flux ($F_{\rm disc}$) to the total spectral flux ($F_{\rm tot}$), \texttt{nthComp} photon index ($\Gamma_{\rm nth}$) and scattered fraction of Comptonized seed photons ($f_{\rm scatt}$) with QPO rms amplitude (${\rm rms}_{\rm QPO}~(\%)$) for four BH-ULXs are depicted from top to bottom panel, respectively. Results from different BH-ULXs are presented using filled points with different styles and colors as marked at the top of the figure. See text for details.}
	\label{fig:QPO_corr}
\end{figure}

\subsection{Spectro-Temporal Correlations}

Next, we examine the correlations among the spectro-temporal properties of the four BH-ULXs. We depict the obtained results in Fig. \ref{fig:QPO_corr}, where the variation of $F_{\rm nth}/F_{\rm tot}$ (in per cent), $F_{\rm disc}/F_{\rm tot}$ (in per cent), $\Gamma_{\rm nth}$, and $f_{\rm scatt}$ with QPO rms amplitudes (${\rm rms}_{\rm QPO}$) are presented from top to bottom panels, respectively. Here, $F_{\rm nth}$ is the \texttt{nthComp} flux, $F_{\rm disc}$ is the \texttt{diskbb} flux, $F_{\rm tot}$ is the total spectral flux, and $f_{\rm scatt}$ denotes the scattered fraction of the seed photons. In each panel, filled points with different styles and colors denote results corresponding to different sources as marked at the top of the figure. We observe that Comptonized flux generally contributes about $50-90\%$, whereas disc flux ranges within $3-50\%$ across the overall range of ${\rm rms}_{\rm QPO}~(\%)$. However, we notice that the flux contributions for NGC 5408 X$-1$ seem coarsely dependent on ${\rm rms}_{\rm QPO}~(\%)$. In addition, $f_{\rm scatt}$ is seen to vary in the range of $2-20\%$ that eventually corroborates the signature of Comptonization of the seed photons. Further, the photon index ($\Gamma_{\rm nth}$) is found to vary in the range of $1.48-2.65$ for all the sources. Note that NGC 6946 X$-1$ and NGC 5408 X$-1$ show relatively softer spectral characteristics ($\Gamma_{\rm nth}\gtrsim 2$) compared to other sources ($\Gamma_{\rm nth}\lesssim 2$) which is consistent with the variation of emission components ($F_{\rm disc}$ and $F_{\rm nth}$) to the total flux ($F_{\rm tot}$) of the sources. 

Overall, all the above findings, as presented in Section 5-6, plausibly indicate that the accretion flow seems to be comprised with a hot Comptonized component resides at the vicinity of the central source, and a relatively low temperature disk confined around the disk equatorial plane. Accordingly, in the next section, we adopt a model of the relativistic, dissipative, low angular momentum accretion flow around the rotating black hole and calculate the global transonic accretion solutions containing shock wave, where hot and dense post-shock flow presumably mimic the Compton corona. Subsequently, we employ the solution of this kind to explain the observed QPO frequency and disc luminosity for a given BH-ULX, and attempt to constrain the physical parameters of the the source, namely the mass, spin and accretion rate.

\section{On the Mass, spin and accretion rate of BH-ULXs}

\begin{figure}
	\begin{center}	
		\includegraphics[width=\columnwidth]{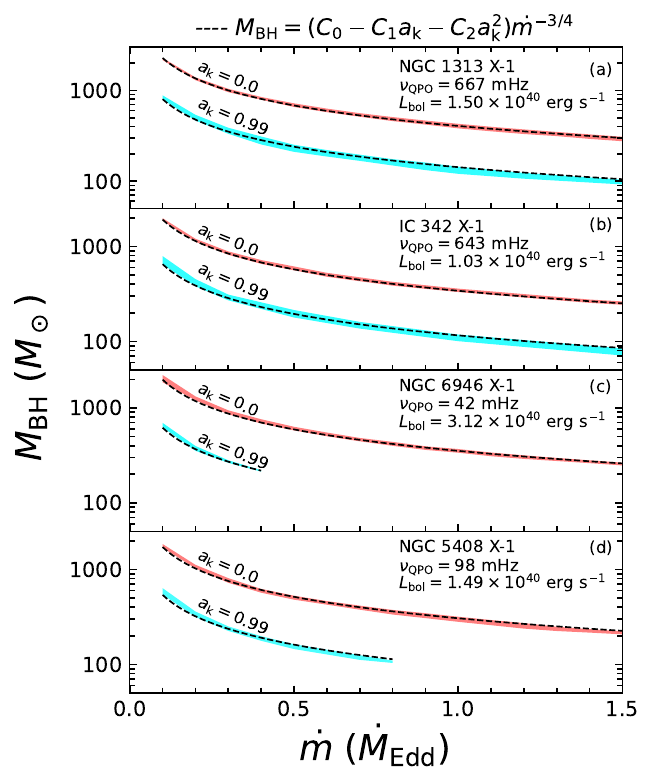}
	\end{center}
	\caption{Correlation between mass accretion rate ($\dot{m}$) and  black hole mass ($M_{\rm BH}$) for different spin parameters of each source. In each panel, the regions shaded with orange and cyan indicate the mass ranges for spin parameter $a_{\rm k}=0.0$ and $0.99$, respectively. The source details and corresponding luminosity and QPO frequency values are also mentioned in each panel. See the text for details.
	}
	\label{fig:mdot-mbh}
\end{figure}

	\begin{table*}
		\renewcommand{\arraystretch}{1.0}
	\caption{Coefficients of the fitted functional form $M_{\rm BH} = (C_{0}-C_{1} a_{\rm k} - C_{2} a_{\rm k}^2)$ $\dot{m}^{-{3/4}}$ along with the black hole mass and accretion rate correlation for spin parameter $a_{\rm k}=0.0$ and $0.99$. The obtained mass and accretion rate ranges are also tabulated for each source.}
	\resizebox{2.1\columnwidth}{!}{%
		\begin{tabular}{l c c c c c c c c c c c}
			\hline\hline
			
			& & & \multicolumn{3}{|c|}{$a_{\rm k}=0$} & & \multicolumn{4}{|c|}{$a_{\rm k}=0.99$}  &  \\
			
			\cline{4-6}
			\cline{8-12} \\
			
			Source & $\nu_{\rm QPO}^{\rm max}$ & $L_{\rm bol}$ & $C_{0}$ & $M_{\rm BH}$  &  Remarks &  & $C_{0}$ & $C_{1}$ & $C_{2}$  & $M_{\rm BH}$  & Remarks \\
			
			& (mHz) & ($10^{40}$ erg $\rm s^{-1}$) &  & ($M_\odot$) & &  &  &  &  & ($M_\odot$)  &  \\ 
			
			\hline
			NGC $1313$ X$-1$ & $667$ & $1.50$ & $405.73$ & $275-2283$ & $\dot{m}\lesseqqgtr \dot{M}_{\rm Edd}$ &  & $352.73$ & $31.83$ & $182.32$ & $92-883$ & $\dot{m}\lesseqqgtr \dot{M}_{\rm Edd}$ \\ \\

			IC $342$ X$-1$ & $643$ & $1.03$ & $340.82$ & $255-1985$ & $\dot{m}\lesseqqgtr \dot{M}_{\rm Edd}$ &  & $336.78$ & $54.72$ & $170.36$ & $71-791$ & $\dot{m}\lesseqqgtr \dot{M}_{\rm Edd}$ \\
			\\
			
			NGC $6946$ X$-1$ & $42$ & $3.12$ & $352.72$ & $248-2221$ & $\dot{m}\lesseqqgtr \dot{M}_{\rm Edd}$ &  & $325.72$ & $37.51$ & $182.24$ & $214-696$ & $\dot{m}<\dot{M}_{\rm Edd}$ \\
			\\
			NGC $5408$ X$-1$ & $98$ & $1.49$ & $305.61$ & $207-1851$ & $\dot{m}\lesseqqgtr \dot{M}_{\rm Edd}$ &  & $315.54$ & $28.62$ & $195.21$ & $104-635$ & $\dot{m}<\dot{M}_{\rm Edd}$ \\

			\hline\hline
		\end{tabular}%
	}
	\label{table:fit_para}
	\hskip -6 cm Note: $M_\odot$ refers the solar mass, and ${\dot M}_{\rm Edd}$ denotes the Eddington accretion rate.
\end{table*}

We utilize the observational findings of the four BH-ULXs under consideration and estimate their possible mass ($M_{\rm BH}$) range in terms of the accretion rate ($\dot m$) and the spin ($a_{\rm k}$) of the source. While doing so, we follow the formalism developed by \cite{Das-etal2021} in order to study the relativistic viscous accretion flow around rotating black hole. In the work of \cite{Das-etal2021}, a model of relativistic, steady, viscous, optically thin, advective accretion flow around a ULX source is developed, where the space-time geometry around the rotating black holes is satisfactorily described using a recently formulated effective potential \cite[]{Dihingia-etal2018}. With this, we solve the flow equations that govern the vertically averaged flow motion in presence of radiative coolings, namely bremsstrahlung, cyclo-synchrotron, and Comptonization processes. While solving the governing equations, we adopt a closer equation in the form of relativistic equation of state to describe a thermally relativistic flow and calculate the shock induced global accretion solutions around the BH-ULXs by means of the input model parameters, namely energy (${\cal E}_{\rm in}$) and angular momentum ($\lambda_{\rm in}$) of the flow at the inner critical point ($r_{\rm in}$), mass accretion rate (${\dot m}$), mass ($M_{\rm BH}$) and spin ($a_{\rm k}$), respectively. Using the shock properties, $i.e.$, shock location ($r_s$) and post-shock velocity ($u$), we pragmatically calculate the QPO frequency as $\nu_{\rm QPO} \sim t^{-1}_{\rm infall}$, where $t_{\rm infall} = \int^{r_i}_{r_s}u^{-1}dr$, $r_i$ being the inner edge of the disk \cite[and references therein]{Molteni-etal1996,Aktar-etal2015,Das-etal2021}. In addition, we also calculate the disk luminosity corresponding to the shocked accretion solutions as $L=4 \pi \int_{r_f}^{r_i} {\cal G} Q r H dr$, where $r_f$ is the outer edge of the disk, $Q$ is the total cooling rates, $H$ is the local disk half-thickness, and ${\cal G} ~(\approx 1-\frac{2r}{r^2 + a^2_{\rm k}})$ denotes the gravitational red-shift factor \cite[]{Das-etal2021}. In this work, total cooling ($Q$) involves bremsstrahlung, cyclo-synchrotron, and comptonization processes \cite[]{Mandal-Chakrabarti2005}
 
Using the input model parameters, namely ${\cal E}_{\rm in}$ and $\lambda_{\rm in}$, we compute the shock induced global accretion solutions for a given set of $({\dot m}, M_{\rm BH}, a_{\rm k})$, that provide the observed $\nu_{\rm QPO}$ and $L_{\rm bol}$ of the BH-ULX sources. Here, we choose viscosity as $\alpha = 0.01$ for the purpose of representation. Subsequently, by tuning ${\cal E}_{\rm in}$ and $\lambda_{\rm in}$, we calculate the maximum observed QPO frequency ($\nu^{\rm max}_{\rm QPO}$, see Table \ref{table:PDS_parameters}) and the corresponding luminosity ($L_{\rm bol}$, see Table \ref{table:Comp_spec_parameters}) of the given BH-ULX source. Here, we consider $\nu^{\rm max}_{\rm QPO}$ as a reference frequency to corroborate the theoretical model prediction. Needless to mention that other observed QPO frequencies can also be explained with the adopted model (see Appendix-C). The obtained results are presented in Fig. \ref{fig:mdot-mbh}, where the correlation between the accretion rate ($\dot m$) and the mass ($M_{\rm BH}$) of a given BH-ULXs is depicted as function of $a_{\rm k}$. In each panel, regions shaded with orange and cyan colors denote the results computed for $a_{\rm k} = 0$ and $0.99$, respectively. Using these results, we obtain the empirical functional form $M_{\rm BH} = (C_{0} - C_{1} a_{\rm k} - C_{2} a_{\rm k}^2)$ $\dot{m}^{-3/4}$, which is found to vary as seemingly exponential with ${\dot m}$ denoted by the dashed curves. Note that the coefficients $C_{0}$, $C_{1}$ and $C_{2}$ are source specific and their values depend on $a_{\rm k}$ as well (see Table \ref{table:fit_para}).

In Fig. \ref{fig:mdot-mbh}a, we depict the results for NGC 1313 X$-1$, where $\nu^{\rm max}_{\rm QPO} = 667$ mHz and $L_{\rm bol}= 1.50 \times 10^{40}$ erg s$^{-1}$ are considered. With this, we find that the source NGC 1313 X$-1$ can accrete at both sub- and super-Eddington accretion rate ($0.1 \le {\dot m} \le 1.5$). Considering its limiting spin values, we compute the mass range of the central source which is obtained as $275 \lesssim \left(M_{\rm BH}/M_{\odot}\right) \lesssim 2283$ for $a_{\rm k} = 0$, and $92 \lesssim \left(M_{\rm BH}/M_{\odot}\right) \lesssim 883$  for $a_{\rm k} = 0.99$, respectively (see Table \ref{table:fit_para}). Furthermore, we also attempt to explain the observed $\nu_{\rm QPO}$ ($\sim 305$ mHz and $\sim 82$ mHz) and the corresponding $L_{\rm bol}$ ($\sim 1.59 \times 10^{40}$ erg s$^{-1}$ and $\sim 1.49 \times 10^{40}$ erg s$^{-1}$) of this source using our theoretical model and present the obtained results in Appendix-C.

We present the results for IC 342 X$-1$ in Fig. \ref{fig:mdot-mbh}b. For this source, we obtain maximum QPO frequency as $\nu^{\rm max}_{\rm QPO} = 643$ mHz and the corresponding luminosity as $L_{\rm bol}=1.03 \times 10^{40}$ erg s$^{-1}$. We observe that the BH-ULX source IC 342 X$-1$ may accrete at both sub- and super-Eddington accretion rate ($\dot m$) irrespective to its spin value ($a_{\rm k}$). Given that the mass of the central source ($M_{\rm BH}$) is initially seen to decrease rapidly with $\dot m$ and then gradually tends to follow its asymptotic value for larger $\dot m$. In particular, for $0.1 \le {\dot m} \le 1.5$, we obtain $255 \lesssim \left(M_{\rm BH}/M_{\odot}\right) \lesssim 1985$ when $a_{\rm k} = 0$, and $71 \lesssim \left(M_{\rm BH}/M_{\odot}\right) \lesssim 791$  for  $a_{\rm k} = 0.99$, respectively (see Table \ref{table:fit_para}).

In case of NGC 6946 X$-1$, we consider $\nu^{\rm max}_{\rm QPO} = 42$ Hz and $L_{\rm bol}= 3.12 \times 10^{40}$ erg s$^{-1}$ and obtain the ${\dot m}-M_{\rm BH}$ correlation by means of the source spin ($a_{\rm k}$). The obtained results are presented in Fig. \ref{fig:mdot-mbh}c, where we observe that the central source accretes at sub-Eddington accretion rate only (${\dot m} \lesssim 0.4$) when $a_{\rm k} \rightarrow 1$. On contrary, the source can accrete at both sub- and super- Eddington accretion rate provided the central source is weakly rotating ($a_{\rm k} \rightarrow 0$). Considering this, we obtain the limiting range of the source mass and the results are obtained as $248 \lesssim \left(M_{\rm BH}/M_{\odot}\right) \lesssim 2221$ when $a_{\rm k} = 0$, and $214 \lesssim \left(M_{\rm BH}/M_{\odot}\right) \lesssim 696$  for  $a_{\rm k} = 0.99$, respectively (see Table \ref{table:fit_para}).

For NGC 5480 X$-1$, we choose $\nu^{\rm max}_{\rm QPO} = 98$ Hz and $L_{\rm bol}=1.49 \times 10^{40}$ erg s$^{-1}$ and compute ${\dot m}-M_{\rm BH}$ correlation in terms of the spin parameter ($a_{\rm k}$). The results are shown in Fig. \ref{fig:mdot-mbh}d, where we find that the central source accretes at sub-Eddington accretion rate (${\dot m} \lesssim 0.8$), provided it is rapidly rotating ($a_{\rm k} \rightarrow 1$). However, if the central source is weakly rotating ($a_{\rm k} \rightarrow 0$), it may accrete at both sub- and super-Eddington accretion rate. For this source, the mass range is calculated as $207 \lesssim \left(M_{\rm BH}/M_{\odot}\right) \lesssim 1851$ when $a_{\rm k} = 0$, and $104 \lesssim \left(M_{\rm BH}/M_{\odot}\right) \lesssim 635$  for  $a_{\rm k} = 0.99$, respectively (see Table \ref{table:fit_para}).

\section{Discussion and Conclusions}

\label{s:Disscussion}

In this work, we carry out a comprehensive spectro-temporal analysis of five BH-ULXs that exhibit QPO features observed with {\it XMM-Newton}. This study allows us to investigate the accretion dynamics of the ULX sources as well as the nature of the central engine.

With this, we examine the short-term variability of the ULXs and find that the fractional variability ($F_{\rm var}$) of $200$ s binned light curves vary in the range of $1.42-27.28$ per cent. Similar variability properties are also observed for IC 342 X$-$1 and NGC 5408 X$-$1 \cite[see also][]{Pintore-etal2014}. Here, we mention that NGC 6946 X$-$1 is found to be the most variable source with $F_{\rm var}\sim27$ per cent. While computing the HID of the ULXs, we find that the count rate generally increases with HID for all sources under consideration expect IC 342 X$-$1, where source intensity is seen to vary significantly although HID varies marginally. 

We perform the power spectral analysis of all five BH-ULXs and confirm the existence of QPO signatures in the frequency range $8.44-643.40$ mHz in all sources \cite[see also][and references therein]{Atapin-etal2019, Fabrika-etal2021}. In addition, for  NGC 5408 X$-1$, we find an additional significant ($\sigma \sim 2.73$) QPO feature of frequency $\sim 98$ mHz appeared simultaneously along with $\sim 38$ mHz QPO frequency (see Table \ref{fig:PDS_QPO} and Fig. \ref{fig:PDS_QPO}). The rms amplitude and significance of the QPOs for all the sources are seen to vary in the range of $4.25-29.05$ per cent and $1.62-9.41\sigma$, respectively (see Table \ref{table:PDS_parameters}). We further mention that all five BH-ULXs are found to manifest QPO features irrespective to the hardness ratio in the range of $0.05-1.40$, although the similar QPO features are generally observed from BH-XRBs in their hard and intermediate states \cite[]{Remillard-etal2006}.

We carry out comprehensive spectral analysis by incorporating several phenomenological and physical models while modelling the combined {\it EPIC-PN} and {\it EPIC-MOS} spectra of {\it XMM-Newton} in $0.3-10$ keV energy range (see section \ref{s:spectral} for details). We find that the phenomenological model combination (model M1) consisting of two disc components (\texttt{diskbb}) along with a Comptonization component (\texttt{simpl}) satisfactorily describes the observed spectra. Indeed, the model M1 elucidates the spectra in terms of relatively hotter as well as softer standard disc components with temperatures varying in the range of $0.75-3.13$ keV and $0.16-0.60$ keV, respectively. The scattered fraction ($f_{\rm scatt}$) is found to vary in a wide range of $1.99-60.64$ per cent (see Table \ref{table:disc_spec_parameters}), that suggests significant Comptonization of the seed photons originated from the disk. Similar model formalism involving standard and slim disks is used to explain the spectral features of NGC 1313 X$-$1 \cite[]{Walton-etal2020}.

Next, we explore a physical model (model M2) by combining \texttt{nthComp} and \texttt{diskbb} components, and it satisfactorily describes the observed spectra of four BH-ULXs. The best fit model spectra of these sources result the ranges of \texttt{nthComp} photon index as $\Gamma_{\rm nth} = 1.48-2.65$, electron temperature as $kT_{\rm e} = 1.62-3.76$ keV, and inner disk temperature $kT_{\rm in} = 0.14_{-0.02}^{+0.03}-0.54_{-0.06}^{+0.15}$ keV, respectively. These findings are consistent with the spectral study ($\Gamma_{\rm nth}\sim2$, $kT_{\rm e}\sim3.5$ keV) of two ULXs in the galaxy NGC 925  \cite[see][]{Pintore-etal2018}. We obtain the optical depth and Compton y-parameter values in the range of $7.65-19.94$ and $1.16-6.24$ that infer the existence of an optically thick medium surrounding the sources. Such a scenario of cool and optically thick accretion disc around ULXs is discussed in \cite{Gladstone-etal2009, Pintore-etal2018,Kobayashi-etal2019}. In addition, we calculate the bolometric luminosity ($L_{\rm bol}$) of four BH-ULXs using model M2 and find that $L_{\rm bol}=(0.59-9.34)\times10^{40}$ erg $\rm s^{-1}$. Note that $L_{\rm bol}$ remains consistent regardless the choice of different model combinations.

We also observe that the correlation between disc luminosity ($L_{\rm disc}$) and color corrected inner disc temperature ($T_{\rm col}$) can be described as $L_{\rm disc} \propto T_{\rm col}^{\alpha}$ with negative exponent ($\alpha$) for all the sources (see Section \ref{s:spec-temp} and Fig. \ref{fig:lum_temp}). These findings are in contrast as the standard accretion disc theory \cite[]{Shakura-etal1973} yields $L_{\rm disc} \propto T_{\rm col}^{4}$. Note that \cite{Kajava-etal2009} reported similar anti-correlation between $L_{\rm disc}$ and $T_{\rm col}$ considering number of ULXs including NGC 5408 X$-1$ and NGC 1313 X$-1$ and obtained similar power-law index ($\alpha \sim -3.5$) as compared to the present results ($\alpha \sim -3.58$). In addition, the source NGC 6946 X$-$1 and NGC 5408 X$-$1 render $\alpha \sim -8.93$ and $-10.31$, respectively (see Fig. \ref{fig:lum_temp}). Similar findings of both positive and negative correlations between $L_{\rm disc}$ and $T_{\rm col}$ for ULXs are reported in \cite{Feng-etal2009} and \cite{Brightman-etal2020}. In explaining the anti-correlation, it is suggested that one can interpret the observed spectra in terms of outflow mechanism. The outflowing matter may enhance the luminosity along the line of sight, resulting an increase in column density also \cite[]{Kajava-etal2009}. However, we find no or very weak correlation (anti-correlation) between column density and luminosity (temperature). We also find positive correlations between $L_{\rm bol}$ and $\Gamma_{\rm nth}$ for all four BH-ULXs that eventually indicates the softening of the observed spectra with the increase of $L_{\rm bol}$. Such correlations are satisfactorily explained using the liner relationship between $L_{\rm bol}$ and $\Gamma_{\rm nth}$ (see Fig. \ref{fig:gamma_lum}) as is generally seen in several BH-XRBs and ULXs \cite[]{Kajava-etal2009}.

Further, we explore the spectro-temporal correlations for all four BH-ULXs. We find that the Comptonized emission dominates ($F_{\rm nth} \sim 50-90 \% F_{\rm tot}$) over the disc emission ($F_{\rm disc} \sim 50 \% F_{\rm tot}$) up to $15\% rms_{\rm QPO}$, whereas both disc and Comptonized emissions contribute almost similarly for $rms_{\rm QPO}> 15\%$. In addition, the scattered fraction ($f_{\rm scatt}$) of the seed photons via Comptonization is found to be in the range of $\sim 10-20\%$ irrespective to $rms_{\rm QPO}$ values except few observations where $f_{\rm scatt} < 10\%$. We also observe that  \texttt{nthComp} belongs to intermediate range as $\sim 1.5-2.3$. Based on the above findings, we infer that the effect of Comptonization \cite[][and references therein for BH sources]{Sreehari-etal2020, Gill-etal2021,Majumder-etal2022,Garcia-etal2022,Zhang-etal2022,Titarchuk-Seifina2023,Peirano-etal2023,Cao-etal2023} plays a viable role in regulating the accretion dynamics around BH-ULXs.

Meanwhile, a number of models have been put forward to explain the nature of the ULXs from theoretical front. \cite{Poutanen-etal2007} suggested a super-critical accretion model for ULXs, accounting the effects of advection and outflow. A good agreement between the derived luminosity-temperature relation with the observations is seen with this model. In addition, the effects of partially inhomogeneous wind in spectro-temporal variability of ULXs is also discussed extensively \cite[]{Middleton-etal2015}. However, the model is found to be dependent on inclination angle and mass accretion rate. Further, \cite{Ambrosi-etal2022} pointed out that the super-Eddington accretion onto black holes of masses in the range of $35-55M_\odot$ with mass ejection scenario  is congruous with the properties of the ULXs. Alternatively, \cite{Narayan-etal2017} proposed a model based on the general relativistic magnetohydrodynamic (GRMHD) simulation for a stellar mass black hole accreting at super-Eddington rate. Indeed, depending on the mass accretion rate, black hole spin, inclination angle and magnetic field configuration, the model yields a luminosity above $\sim 10^{40}$ erg $\rm s^{-1}$. In addition, \cite{Mondal-etal2019} proposed an advective magnetized accretion-ejection mechanism around a stellar mass black hole that explains the observational aspects of ULXs in power-law dominated state. Recently, \cite{Das-etal2021} developed a physically motivated model by incorporating the relativistic dissipative accretion flow in ULXs. Adopting this model formalism, it is found that accretion in both massive stellar-mass and intermediate mass black holes can power the ULX sources, depending on their mass accretion rate and the spin values (see Fig. \ref{fig:mdot-mbh} and Table \ref{table:fit_para}). In particular, we observe that NGC 6946 X$-1$ and NGC 5408 X$-1$ possibly accrete at sub-Eddington rate provided their central sources are rapidly rotating ($a_{\rm k} \rightarrow 1$). On contrary, IC 342 X$-1$ and NGC 1313 X$-1$ seems to accrete either at sub- or  super-Eddington accretion rate depending on the spin values ($0\le a_{\rm k}\le 0.99$) of the central accretor.

Finally, we mention the limitation of the present work. In our analysis, we consider that the ULX sources under consideration harbour black hole at their central core. However, the nature of the central source in ULXs could be neutron star as well even with no detection of pulsed emission \cite[]{King-etal2017,Walton-etal2018b}. Hence, we indicate that the overall conclusions delineated in this work are subjected to alter for neutron star accretors as the inner boundary conditions for neutron star accretion differ from black hole system \cite[]{Frank-etal2002}. This aspect is indeed relevant, however its implementation is beyond the scope of the present work, which we plan to take up as future work.

\section*{Acknowledgments}

Authors thank the anonymous reviewer for constructive comments and useful suggestions
that help to improve the quality of the manuscript. SD thanks Science and Engineering Research Board (SERB), India for support under grant MTR/2020/000331. SM and SD also thank the Department of Physics, IIT Guwahati, India for providing the facilities to complete this work. VKA and AN thank GH, SAG; DD, PDMSA, and Director, URSC for encouragement and continuous support to carry out this research. This publication uses the data from the {\it XMM-Newton} mission, archived at the HEASARC data centre. The instrument team is thanked for processing and providing the useful data as well as software for this analysis.

\section*{Data Availability}

Data used for this publication are currently available at the HEASARC browse website (\url{https://heasarc.gsfc.nasa.gov/db-perl/W3Browse/w3browse.pl}).

\appendix

\section {Estimation of ${\rm QPO}_{\rm rms}\%$ for M82 X$-$1}

	Since M82 X$-$2 is extremely variable \cite[]{Brightman-etal2016b}  and M82 X$-$1 is located within the 5 arcsec separation of it, the effect of possible contamination can not be ruled out. This causes the spectral analysis of M82 X$-$1 challenging as the obtained results could be significantly biased and hence, we refrain from spectral study of M82 X$-$1. However, we estimate ${\rm QPO}_{\rm rms}\%$ as the QPO feature originates from M82 X$-$1 only. While estimating $QPO_{\rm rms}\%$, we make use of the nearby {\it Chandra} observations of M82 X-1 and M82 X-2, where we consider $3$ arcsec circular region about each source and obtain the count rates of individual sources. Further, we consider $8$ arcsec circular region at the center which includes both sources (M82 X-1 and M82 X-2). Evidently, this provides the total count rates obtained from the contribution of both sources. Accordingly, the photon fraction for each source is calculated as the ratio of individual source counts to the total counts \cite[see also][]{Feng-Kaaret2007,Brightman-etal2019}. We divide the estimated rms with this photon fraction of M82 X-1 and express the corrected rms in percentage. Note that \cite{Feng-Kaaret2007} obtained the photon fraction from the `surface brightness fit of {\it XMM} Image' and in this study, we estimate ${\rm QPO}_{\rm rms}\%$ using the near by {\it Chandra} observations. The details of {\it XMM} and {\it Chandra} observations \cite[]{Brightman-etal2019} along with the estimated ${\rm QPO}_{\rm rms}\%$ are tabulated in Table A1. 

\begin{table*}
	\centering
	\renewcommand{\arraystretch}{1}
	\vspace*{-0.39cm}  
	\hspace*{+1cm}
	\caption{Details of M82 X$-1$ observations with {\it XMM-Newton} and {\it Chandra}. Corrected ${\rm QPO}_{\rm rms}\%$ and Total$_{\rm rms}\%$ are presented in column 7 and 9. See text for details.}
	\label{tab1}
	\addtolength{\tabcolsep}{-3pt}  
	\resizebox{\textwidth}{!}{%
		\begin{tabular}{l @{\hspace{0.5cm}} c @{\hspace{0.5cm}} c @{\hspace{0.5cm}} c @{\hspace{0.3cm}} c @{\hspace{0.3cm}} c @{\hspace{0.3cm}} c @{\hspace{0.5cm}} c @{\hspace{0.4cm}} c} 
			\hline
			{\it XMM} & QPO Freq. & Estimated  & {\it Chandra} & Date & M82 X$-1$ & Corrected & Estimated &  Corrected\\
			
			 Epoch & $(\nu_{\rm QPO})$ & $\rm QPO_{\rm rms}(\%)$ & ObsID & of Obs. &  Contribution$^*$($\%$) & $\rm QPO_{\rm rms}(\%)$$^\dagger$ & $\rm Total_{\rm rms}$ (\%) & $\rm Total_{\rm rms}$ (\%)$^\dagger$\\	
			
			\hline

			XM1 & $55.85_{-2.01}^{+1.44}$ & $4.25\pm1.02$ & 02933 & 2002-06-18 & 36.84 & 11.54$\pm4.33$ & $12.04\pm1.93$ & $32.68\pm10.30$ \\
			
			XM2  & $110.81_{-4.26}^{+5.03}$ & $8.35\pm1.15$ & 06097 & 2005-02-04 & 68.84 & 12.13$\pm2.18$ & $13.91\pm2.33$ & $20.21\pm4.06$ \\
			
			XM5 & $45.55_{-2.19}^{+1.61}$ & $5.97\pm1.12$ & 10545 & 2010-07-28 & 27.28 &  21.88$\pm 7.75$ & $18.68\pm2.28$ & $68.48\pm12.84$ \\
			
			XM6 & $47.63_{-3.22}^{+5.47}$ & $8.99\pm1.42$ & 10545 & 2010-07-28 & 27.28 &  32.95$\pm 11.18$ & $17.63\pm4.25$ & $64.62\pm24.32$ \\
			
			XM7 & $35.74_{-2.44}^{+2.53}$ & $7.72\pm1.74$ & 10545 & 2010-07-28 & 27.28 &  28.30$\pm10.64$ & $11.66\pm2.02$ & $42.74\pm14.28$ \\
			
			\hline
			
		\end{tabular}%
	}
		\begin{list}{}{}
		\item $^{*}$Contribution of M82 X$-1$ in the total count rate within 8 arcsec {\it Chandra} aperture. 
		\item $^\dagger$${\rm QPO}_{\rm rms}\%$ values involve uncertainty and hence should be referred accordingly.
	\end{list}
	
	\addtolength{\tabcolsep}{3pt}
\end{table*}

\section{Flux and temperature correlation }

We examine whether the apparent correlation between $L_{\rm disc}$ and $T_{\rm col}$ is actually a result of degeneracy or not. In doing so, we calculate disc luminosity as $L_{\rm disc} = 4\pi d^2 F_{\rm disc}$, where $F_{\rm disc}$ is obtained from the spectral fitting with convolution model \texttt{cflux}. We compute $T_{\rm col}$ from the inner disc temperature $T_{\rm in}$ by multiplying a constant spectral hardening factor ($f$). Therefore, it is appropriate to perform degeneracy check between $F_{\rm disc}$ and $T_{\rm in}$ instead of $L_{\rm disc}$ and $T_{\rm col}$. We run an MCMC simulation using the Goodman-Weare algorithm \cite[]{Goodman-etal2010} with a chain length of $20000$. The obtained results are shown in Fig. B1 for all the sources under consideration except M82 X$-$1. Note that $F_{\rm disc}$ depends on $T_{\rm in}$ for each source (ruling out the degeneracy between the parameters) which possibly causes the observed correlation between $L_{\rm disc}$ and $T_{\rm col}$ for the ULXs under consideration except M82 X$-$1.

\begin{figure*}
		\subfloat[IC 342 X$-1$]{\includegraphics[clip = true, width=0.45\textwidth, height=6.75cm]{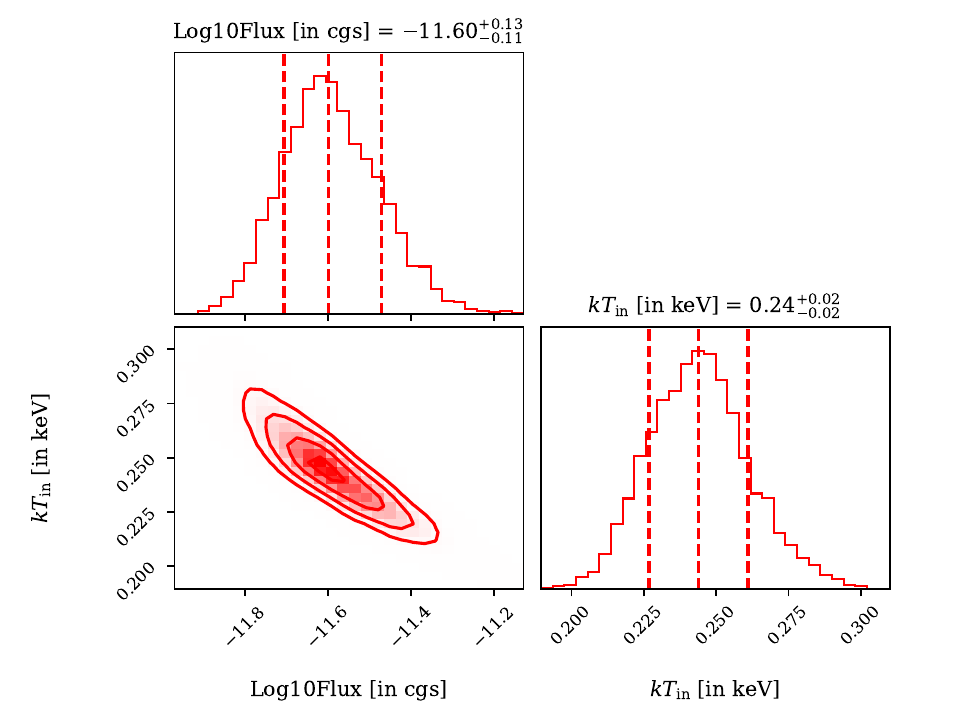}}
		\subfloat[NGC 5408 X$-1$]{\includegraphics[clip = true, width=0.45\textwidth, height=6.75cm]{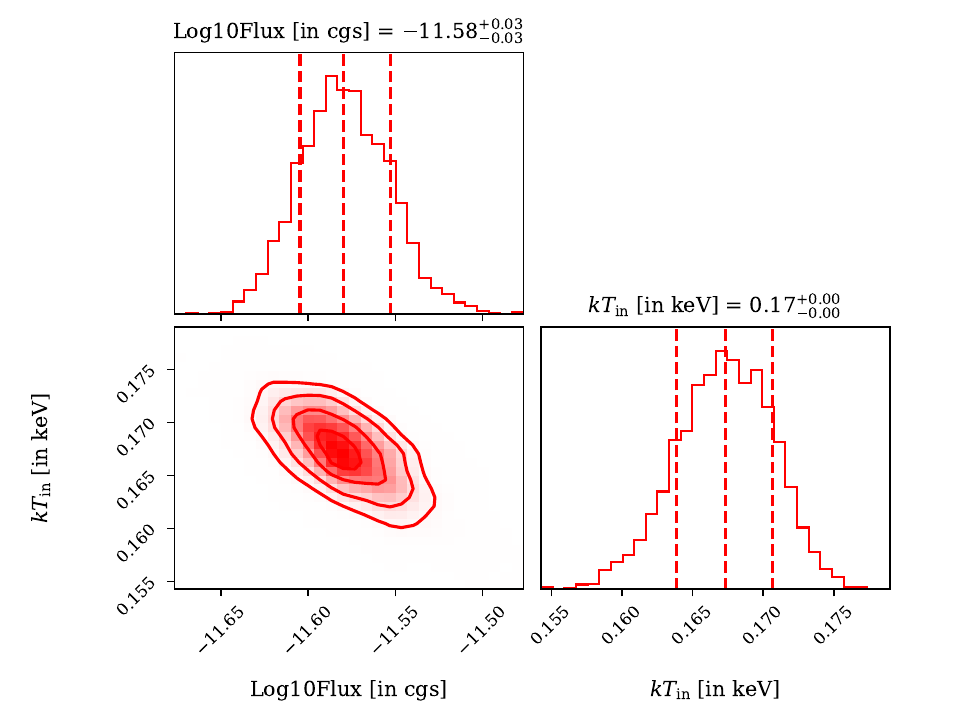}}\\
		\subfloat[NGC 6946 X$-1$]{\includegraphics[clip = true, width=0.45\textwidth, height=6.75cm]{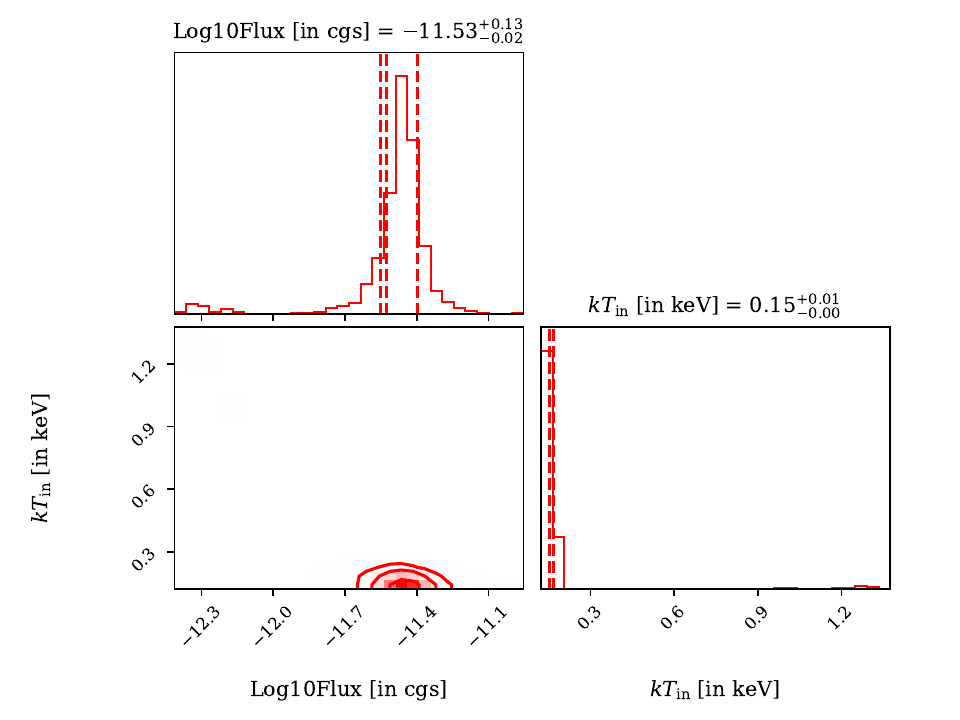}}
		\subfloat[NGC 1313 X$-1$]{\includegraphics[clip = true, width=0.45\textwidth, height=6.75cm]{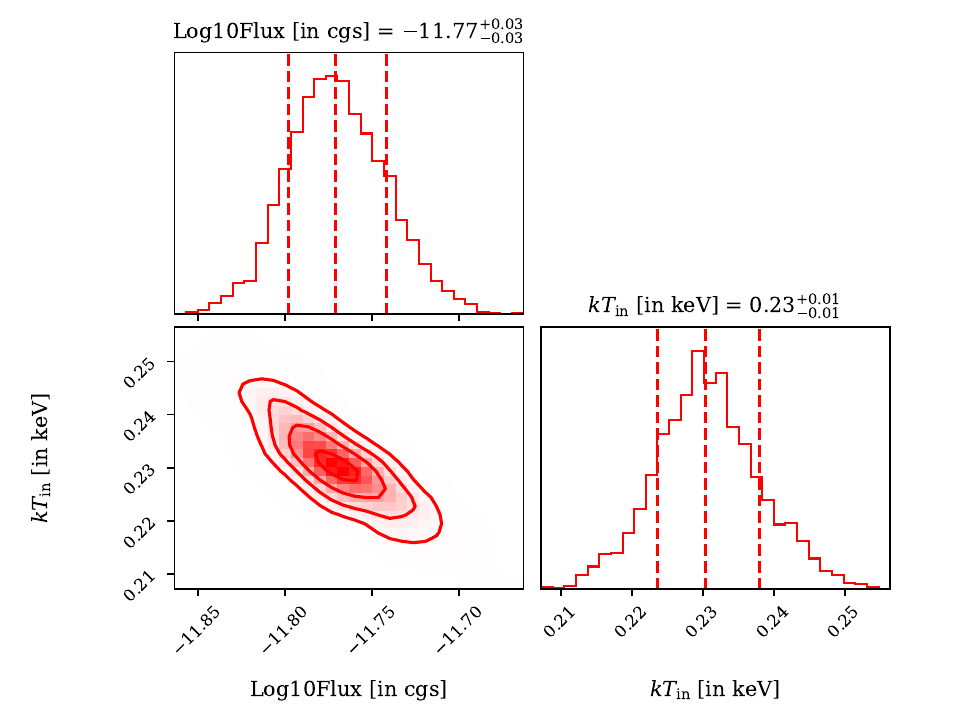}}\\
		\caption{Correlation between the inner disc temperature and estimated bolometric disc flux of each source. The MCMC simulation carried out using the Goodman-Weare algorithm with a chain length of $20000$. The contours in each panel represent $1\sigma$, $2\sigma$, and $3\sigma$ confidence ranges.}
		\label{fig:r1}
	\end{figure*}
	
\section{Dependence of $\nu_{\rm QPO}$ on accretion rate}

\begin{figure}
	
	\includegraphics[width=\columnwidth]{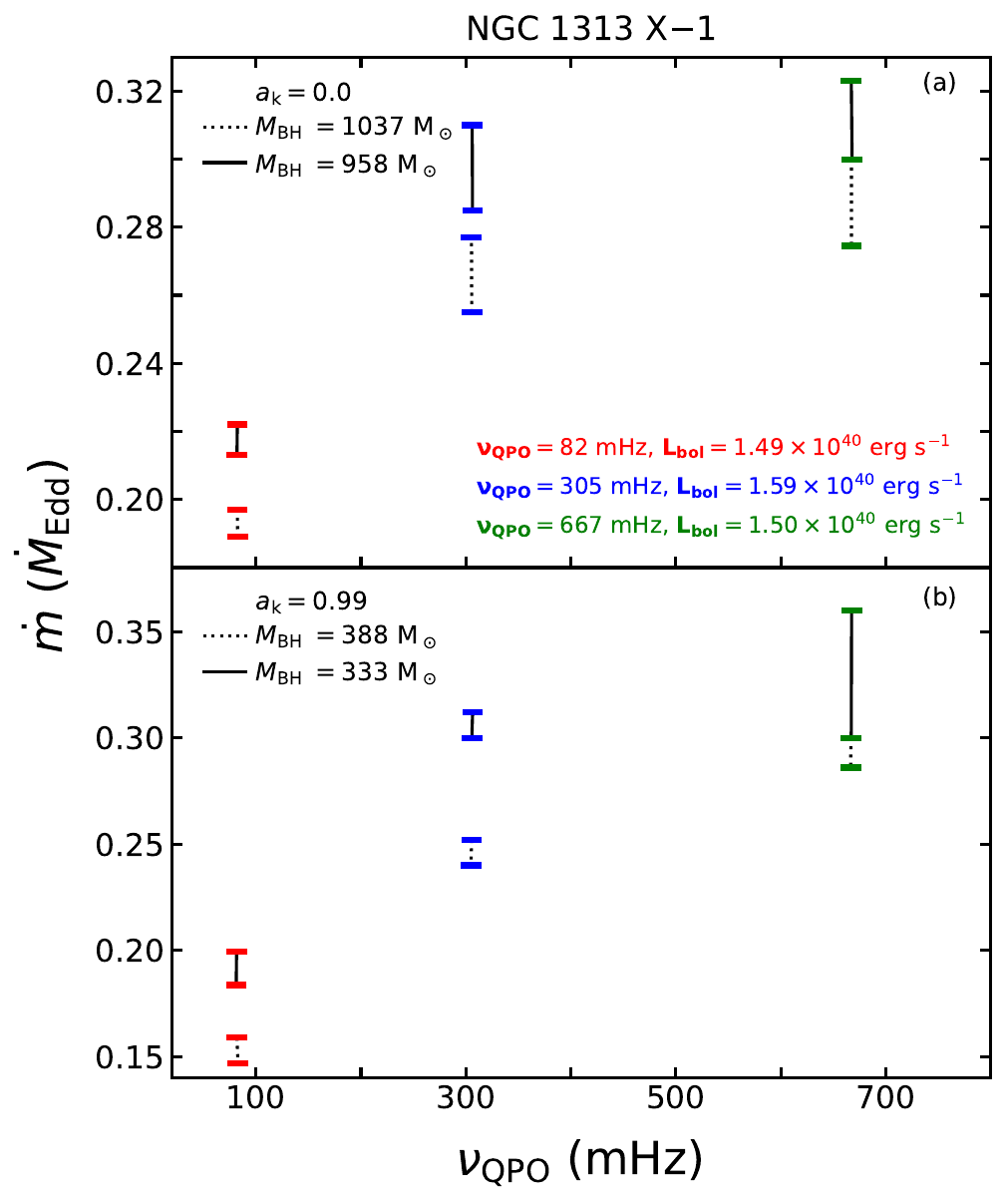}
	\caption{Variation of observed QPO frequency ($\nu_{\rm QPO}$) with model predicted accretion rate ($\dot m$) for NGC 1313 X$-1$. Results corresponding to $a_{\rm k} = 0.0$ and $0.99$ are presented in panel (a) and (b), respectively. See text for details.
	}
	\label{fig:appendixC}
\end{figure}

In order to examine how $\nu_{\rm QPO}$ and $L_{\rm bol}$ of NGC 1313 X$-1$ depends on the accretion rate ($\dot m$), we consider the mass range of the source as $958 \lesssim M_{\rm BH}/M_\odot \lesssim 1037$ for $a_{\rm k} = 0.0$ and $333 \lesssim M_{\rm BH}/M_\odot \lesssim 388$ for $a_{\rm k} = 0.99$ (see Fig. \ref{fig:mdot-mbh}a). With this, we estimate the mass accretion rate in the sub-Eddington regime that satisfactorily explains both $\nu_{\rm QPO}$ and $L_{\rm bol}$. The obtained results are shown in Fig. \ref{fig:appendixC}. In upper (lower) panel, we choose $a_{\rm k}=0.0$ ($0.99$), and solid and dashed vertical lines are for $M_{\rm BH} = 958~M_\odot$ and $1037~M_\odot$ ($M_{\rm BH} = 333~M_\odot$ and $388~M_\odot$). In both panels, the vertical lines flanked by red, blue and green horizontal bars denote results corresponding to ($\nu_{\rm QPO}, L_{\rm bol}) = (82~{\rm mHz}, 1.49 \times 10^{40}~{\rm erg~s}^{-1})$, ($305~{\rm mHz}, 1.59 \times 10^{40}~{\rm erg~s}^{-1}$) and ($667~{\rm mHz}, 1.50 \times 10^{40}~{\rm erg~s}^{-1}$), respectively. It is evident that $\nu_{\rm QPO}$ increases with $\dot m$. We further infer that similar variation of $\nu_{\rm QPO}$ with $\dot m$ is also very much likely in super-Eddington accretion regime as well.

\end{document}